\documentclass[a4paper,oneside,10pt]{article}
\usepackage[ansinew]{inputenc}
\usepackage[T1]{fontenc}
\usepackage{geometry}
\usepackage{amsmath}
\usepackage{amsthm}
\usepackage{url}
\usepackage{graphicx}
\usepackage{graphics}
\usepackage{epstopdf}
\usepackage{multirow}
\usepackage{array}
\usepackage{lmodern}
\usepackage{a4wide}
\usepackage{srcltx}

\begin{document}

\title{Statistical estimation of mechanical parameters of clarinet reeds using experimental and numerical approaches}

\author{Pierre-André Taillard$^1$, Franck Laloë$^2$, Michel Gross$^3$,\\
 Jean-Pierre Dalmont$^4$ and Jean Kergomard$^5$\\
$^1$ Conservatoire de musique neuchâtelois,\\
Avenue Léopold-Robert 34; CH-2300 La Chaux-de-Fonds; Switzerland. \\
$^2$ Laboratoire Kastler Brossel – UMR 8552 Ecole Normale Supérieure,\\
UPMC, CNRS, 24 rue Lhomond; F-75231 Paris Cedex 05 ; France. \\
$^3$ Laboratoire Charles Coulomb  - UMR 5221 CNRS-UM2, Université Montpellier, \\II place Eugène Bataillon;
F-34095 Montpellier ; France. \\
$^4$ Laboratoire d’Acoustique de l’Université du Maine - UMR CNRS 6613,\\ Université du Maine, F-72085 Le
Mans; France. \\
$^5$ Laboratoire de Mécanique et d’Acoustique - CNRS, UPR 7051,\\ Aix-Marseille Univ, Centrale Marseille; F-13402 Marseille Cedex 20;
France.
}
\maketitle

\begin{abstract}
A set of 55 clarinet reeds is observed by holography, collecting 2 series of measurements made under 2 different moisture contents, from which the resonance frequencies of the 15 first modes are deduced. A statistical analysis of the results reveals good correlations, but also significant differences between both series. Within a given series, flexural modes are not strongly correlated. A Principal Component Analysis (PCA) shows that the measurements of each series can be described with 3 factors capturing more than $90\%$ of the variance: the first is linked with transverse modes, the second with flexural modes of high order and the third with the first flexural mode. A forth factor is necessary to take into account the individual sensitivity to moisture content.
Numerical 3D simulations are conducted by Finite Element Method, based on a given reed shape and an orthotropic model. A sensitivity analysis revels that, besides the density, the theoretical frequencies depend mainly on 2 parameters: $E_L$ and $G_{LT}$. An approximate analytical formula is proposed to calculate the resonance frequencies as a function of these 2 parameters.
The discrepancy between the observed frequencies and those calculated with the analytical formula suggests that the elastic moduli of the measured reeds are frequency dependent. A viscoelastic model is then developed, whose parameters are computed as a linear combination from 4 orthogonal components, using a standard least squares fitting procedure and leading to an objective characterization of the material properties of the cane \textit{Arundo donax}.
\end{abstract}




\section{Introduction}

\bigskip Clarinettists experience every day the crucial importance of clarinet reeds for the quality of sound. Their characterization is a real challenge for musicians who wish to obtain reeds that are suited to their personal needs.
The present paper address this complex field of research. Its scope is restricted to the development of an objective method for a mechanical characterization of single reeds of clarinet type.
>From the shape and the resonance frequencies of each individual reed (measured with heterodyne holography), we intend to deduce the mechanical properties of the material composing it. A subsequent study should then examine how these mechanical properties are correlated with the musical properties of the reeds.

Generally, the physicist chooses a model in order to validate it by observations. In the present study, the complexity of the problematic forced us to adopt the reverse attitude: We observe the mechanical behavior of clarinet reeds with a statistically representative sample and exploit afterward the statistical results for establishing a satisfactory mechanical model designed with a minimal number of parameters.

Natural materials, as wood or cane, are often orthotropic and exhibit a different stiffness along the grain (longitudinally) as in the others directions.
The problem is then obviously multidimensional. Nevertheless, reed makers classify their reeds by a single parameter: the nominal reed "strength" (also called "hardness"), in general from 1 to 5, which basically reflects the stiffness of the material (cane, \emph{Arundo donax L.}), since all reeds of the same model have theoretically the same shape. The method of measurement is generally not publicized by manufacturers, but this
"strength" is probably related to the static Young modulus in the longitudinal direction $E_L$.

"Static" (i.e. low frequency) measurements of the elastic parameters of cane are available in the literature, for instance Spatz \emph{et al.} \cite{spatz1997biomechanics}. A viscoelastic behavior has been reported in experimental situations (see e.g. Marandas \emph{et al.} \cite{marandas1994caractérisation}, Ollivier \cite{Ollivier_2002} or Dalmont \emph{et al.} \cite
{dalmont2003nonlinear}) and this fact seems generally well accepted in wood sciences and biomechanics (for instance Speck \emph{et al.} \cite%
{speck2003mechanical,speck2004damped}). Marandas \emph{et al.} proposed a
viscoplastic model of the wet reed. Viscoelastic behavior for cane was already demonstrated by Chevaux \cite%
{Chevaux_1997}, Obataya \emph{et al.} \cite{obataya1996_physical,obataya1999a_effects,obataya1999b_mechanical,obataya1999c_acoustic}
and Lord \cite{lord2003viscoelasticity}. These authors study only the viscoelasticity of the longitudinal Young modulus $E_{L}$, leaving aside the case
of the shear modulus in the longitudinal/tangential plane $G_{LT}$. Furthermore, they give no really representative statistics
about the variability of the measured parameters.

The observation of mechanical resonance frequencies can be achieved by different methods.
The methods used by Chevaux, Obataya and Lord are destructive for the reed, which cannot be used for
further musical tests. On the contrary, holography is a convenient
non-destructive method, the reed
being excited by a loudspeaker. For instance Pinard \emph{et al.} \cite%
{pinard2003musical} measured with this method the frequency of the 4 lowest resonances and
focused their attention on the musical properties of the reeds.

The digital Fresnel holography method was used by Picart \emph{et al.} \cite%
{picart2007tracking,picart2010study} and Mounier \emph{et al.} \cite%
{mounier2008investigation} to measure high amplitude motion of a reed blown
by an artificial mouth. Guimezanes \cite{Guimezanes_2007} used a scanning vibrometer.

Recent technological developments provide very efficient and convenient
measurements with holography, without having to manually identify the modes
of resonance and to be satisfied with a single picture of their vibration: in a
few minutes hundreds of holograms are acquired showing the response of a
reed for many frequencies. The temperature and the moisture content can be
considered as constant during a measurement series\footnote{The significantly lower correlations between resonance frequencies (compared to our data) shows that it was probably not the case in Pinard's study. This fact may also reflect an unprecise determination of the resonance frequencies.}.
The Sideband Digital Holography technique provides additional facilities (see \ref{HoloSetup}).

Different authors (among them Casadonte \cite{casadonte1993perfect,casadonte1995}, Facchinetti \emph{et al.} \cite%
{facchinetti2000application,facchinetti2003numerical} and Guimezanes
\cite{Guimezanes_2007}) modeled the clarinet reed by Finite Elements Method (FEM) and computed the first few eigenmodes. They
chose appropriated values of the elastic parameters in the literature, ignoring however viscoelastic behavior. The goodness of fit between observations and model was of secondary importance, except for Guimezanes. This latter author built a 2-D elastic model of the reed
with longitudinally varying parameters. He fitted his model quite adequately
with his observations (only 5 resonances were measured), but the fitted
parameters seem not really plausible physically. His model didn't respect
the assumption of a radial monotonically decrease of stiffness from the
outer side to the inner side of the cane. Under such conditions, the
frequency of the first resonance would increase in comparison to homogeneous
material, and not decreased, as observed experimentally.

In Section \ref{SDH} the measurement method is presented. The experimental setup is described in Section \ref{ExperiSetup} and the method for observing resonance frequencies is detailed in \ref{ObsMethod}. The results for 55 reeds are
given in Section \ref{ModalFreq} (statistics, Principal Component Analysis (PCA)\cite{jolliffe2002principal}).

In Section \ref{NuSi}, the development and the selection of a satisfactory mechanical model with minimal structure is described. First, a numerical analysis of the resonance frequencies of a reed assumed
to be perfectly elastic is done by Finite Element Method (FEM), and a metamodel computing the resonance
frequencies from elastic parameters is given in Section \ref{section_III_A}. This allows  solving the inverse problem in a fast way. However, because the elastic model is not very satisfactory, viscoelasticity has to be introduced and some parameters are added to the
model in Section \ref{ViMo}. The viscoelastic model has however too many degrees of freedom, according to PCA. Consequently, the viscoelastic parameters of the model are assumed to be correlated and PCA indicates that these parameters can be probably reconstructed from 4 orthogonal components,
as a linear combination, by multiple regression (Section \ref{invProb}).
The relationships between the components and the
viscoelastic parameters is given, and finally the resulting values for these
parameters are discussed in Section \ref{results} and compared with the results of the literature.


\section{Observations by Sideband Digital Holography\label{SDH}}

\subsection{Experimental setup\label{ExperiSetup}}

\subsubsection{Holographic setup\label{HoloSetup}}

\begin{figure}[tbp]
\par
\begin{center}
\includegraphics[width=10cm,keepaspectratio=true]{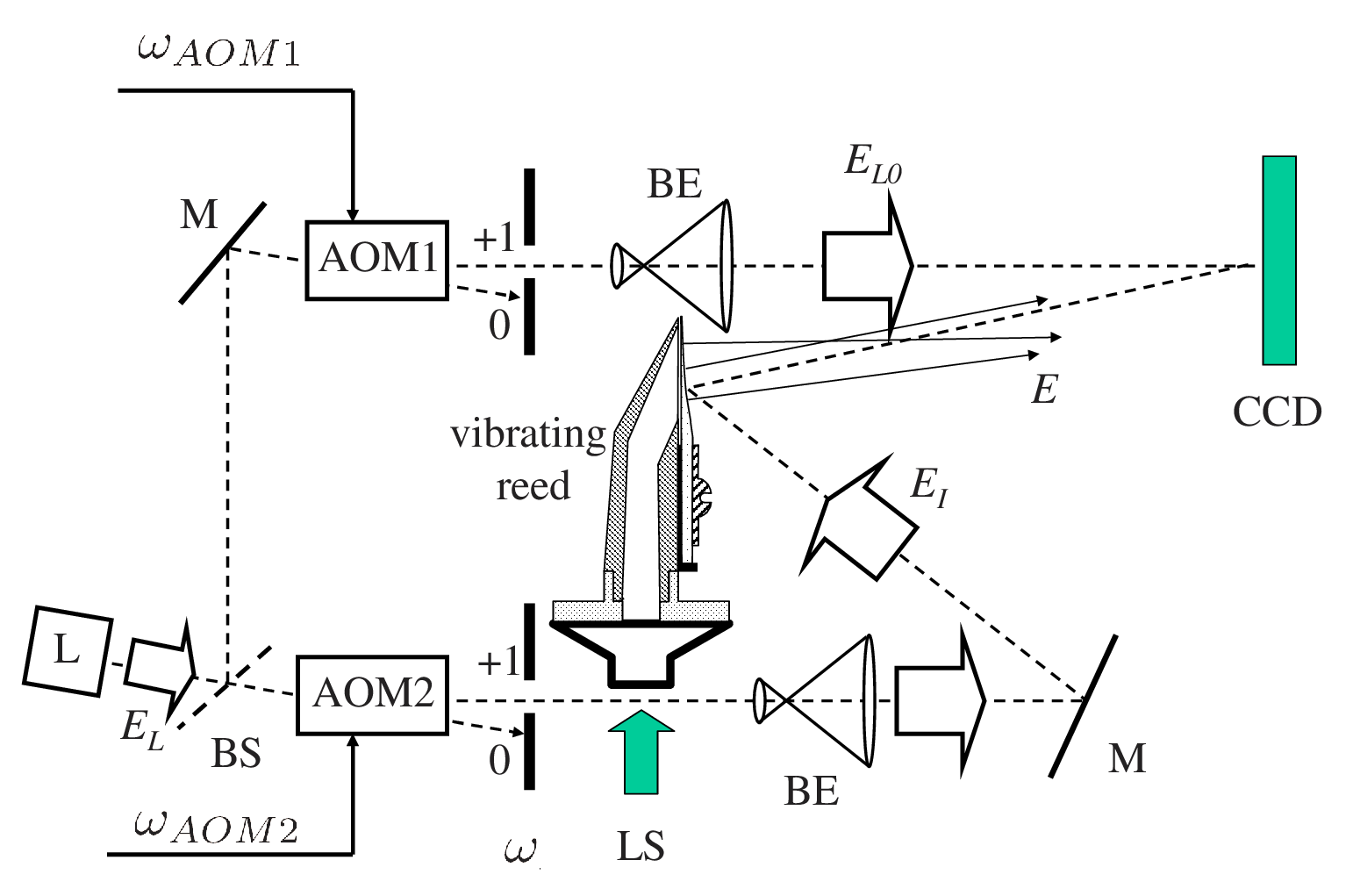}
\end{center}
\caption{ Holographic setup. L: main laser; AOM1, AOM2: acousto-optic
modulators; M: mirror; BS: beam splitter; BE: beam expander; CCD: CCD
camera; LS: loudspeaker exciting the clarinet reed through the bore of a clarinet mouthpiece at frequency $f=%
\protect\omega /2\protect\pi $.}
\label{fig_setup}
\end{figure}

\bigskip
The experimental setup is shown schematically in Fig. \ref{fig_setup}. A
laser beam, with wavelength $\lambda =650$ nm (angular frequency $\omega
_{L} $) is split into a local oscillator beam (optical field $E_{LO}$) and
an illumination beam ($E_{I}$); their angular frequencies $\omega _{LO}$ and
$\omega _{I}$ are tuned by using two acousto-optic modulators (Bragg cells
with a selection of the first order diffraction beam) AOM1 and AOM2: $\omega
_{LO}=\omega _{L}+\omega _{AOM1}$ and $\omega _{I}=\omega _{L}+\omega
_{AOM2} $, where $\omega _{AOM1,2}\simeq 2\pi \times 80$ MHz. The first beam
(LO) is directed via a beam expander onto a CCD camera, while the second
beam (I) is expanded over the surface of the reed, which vibrates at
frequency $f$ . The light reflected by the reed (field $E$) is directed
toward the CCD camera in order to interfere with the LO beam ($E_{LO}$). 4 phases were used (phase shifting digital holography) and we select the first
sideband of the vibrating reed reflected light by adjusting $\omega
_{AOM1,2} $ to fulfil the condition: $\omega _{AOM1}-\omega _{AOM2}=2\pi
(f+f_{CCD})/4$, where $f_{CCD} $ is the CCD camera frame frequency. The
complex hologram signal $H$ provided by each pixel of the camera, which is
proportional to the sideband frequency component of local complex field $E$,
is obtained by 4-phases demodulation: $H=(I_{0}-I_{2})+j(I_{1}-I_{3})$ where
$I_{0}\ldots I_{3}$ are 4 consecutive intensity images digitally recorded by
the CCD camera, and $j^{2}=-1$. From the complex hologram $H$, images of the
reed vibration are reconstructed by a standard Fourier holographic
reconstruction calculation \cite{schnars1994direct}. These holographic
reconstructed images exhibit bright and dark interference fringes. Counting these
fringes provides the amplitude of vibration of the object (in the
direction of the beam), which depends on the wavelength $\lambda $ of the
laser, and on the first Bessel function $J_{1}$, for instance $\pm 95$nm for the first, $\pm 770$nm for the $5^{th}$ and  $\pm 1.6\mu m$ for the $10^{th}$ maximum (bright fringes) \cite%
{joud2009a_imaging,joud2009bfringe}\footnote{%
The original notation from the cited paper is kept. This notation is
only valid for this paragraph.}.

This method has 3 main advantages:

(i) The time for data acquisition is very short, about 3 minutes for
recording 184 holograms, including holographic reconstruction.

(ii) The signal to noise ratio is significantly better than with traditional
technology, particularly through the elimination of signal at zero frequency.

(iii) The visualization of large-amplitude vibration (order of magnitude:
0.1 mm) is possible by using high harmonics orders (up to several hundred
times the excitation frequency).

\subsubsection{Reed excitation}

The reed was excited by a tweeter loudspeaker screwed onto an aluminium
plate, connected to a clarinet mouthpiece. The lay of this mouthpiece was
modified to be strictly flat. A plastic wedge of uniform
thickness has been inserted between the lay and the reed, longitudinally to the same height as
the ligature (Vandoren Optimum), allowing free vibrations of the entire vamp
(length: about 38 mm), see Fig. \ref{fig_setup}. This ensures precise boundary conditions, avoiding
any dependence to deformations of the reed. The repeatability of the
longitudinal placing of the wedge and of the reed was ensured by a
Claripatch ring \cite{Claripatch}.

This setup requires some comments:

(i) The reed is excited exclusively through the bore of the mouthpiece.

(ii) The pressure field in the chamber of the mouthpiece was not measured.
Like for a real instrument, the edges of the reed (protected by the walls of
the chamber) are subject to a pressure field, which is probably lower than
the pressure acting on the rest of the vamp.

(iii) The boundary conditions are very different from those of a real
instrument (no curved lay, no contact with the lip). In addition, the reed
was not moistened for the measurement.

(iv) The excitation device is almost closed. The acoustical resonances of the excitation device are
unknown, but may quite easily be deduced by comparing different measurements, because they are always present at the same frequency.

\subsection{Observation of resonance frequencies}\label{ObsMethod}

\subsubsection{Experimental protocol}

\begin{figure}[tbp]
\begin{center}
\includegraphics[width=14cm,keepaspectratio=true]{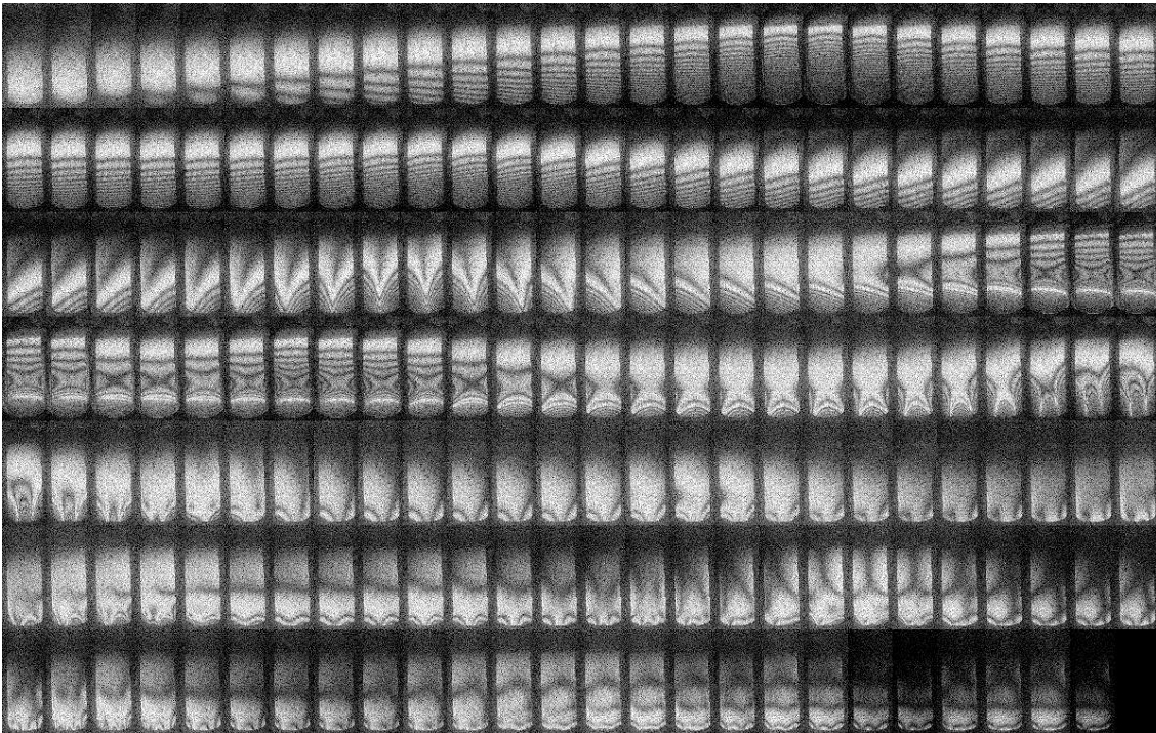}
\end{center}
\caption{ Typical holographic patterns of series A (asymmetrical sinusoidal
excitation: the left side of the reed is more strongly excited than the
right side). Frequency range: 1.4 to 20 kHz by steps of 25 cents (181
pictures ordered from left to right, continued on the next row; the tip of
the reed is down on each picture). Some modes are easily identified: $F1$
(1st row, 19th picture), $T1$ (3rd row, 10th picture), $F2$ (4th row, 7th
picture), $T2$ (4th row, penultimate picture), $X1$ (5th row, 5th picture) etc ...
Modes $T3$ and $F3$ are almost at the same frequency (6th row, 4th and 5th
pictures, probably). The last picture of the 3rd row corresponds to an
acoustic resonance of the excitation device. It is present on all holograms
of both series at the same frequency (examine Fig.\protect\ref{fig_fig2b}).
The excitation amplitude exponentially increases
until the 73rd picture
(3rd line, 21th picture), being hold constant beyond. }
\label{fig_fig2a}
\end{figure}

\begin{figure}[tbp]
\begin{center}
\includegraphics[width=14cm,keepaspectratio=true]{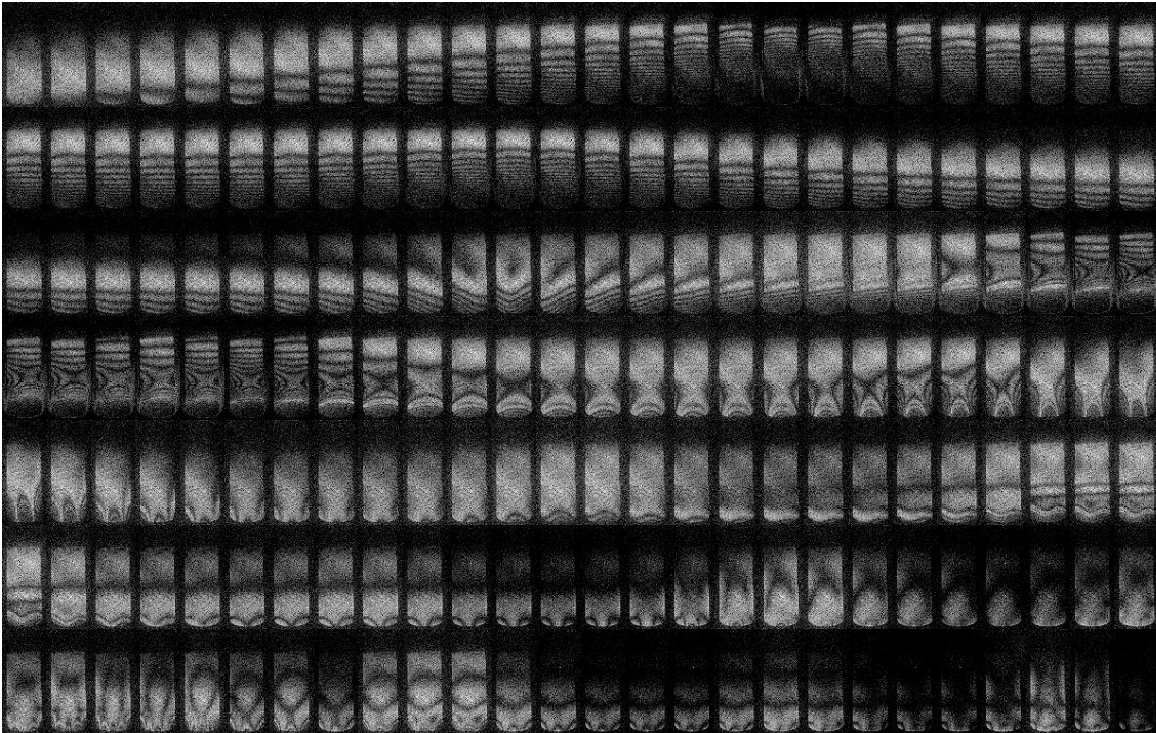}
\end{center}
\caption{ Typical holographic patterns of series B (symmetrical sinusoidal
excitation). The modes $X1$ and $T3$ cannot be distinguished anymore. Notice
that $T1$ is less marked than under asymmetrical excitation (for some reeds
even difficult to identify) and that the pattern has a significant flexural
component, strongly dependent of the lateral placing of the reed on the
mouthpiece. Notice that the symmetry of the patterns near $T1$ depends on
the excitation frequency. }
\label{fig_fig2b}
\end{figure}

55 clarinet reeds of model Vandoren V12 were purchased in a music shop: 12,
12, 20 and 11 reeds of nominal forces 3, $3\frac{1}{2}$, 4 and $4\frac{1}{2}$%
, respectively. 29 reeds were used for two preliminary studies in order to
develop the measurement protocol. Each of these reeds was played a total of
some tens of minutes, spread over several weeks before measurement with the
final protocol. The other 26 reeds were strictly new by measurement, which
was performed immediately after package opening (for 21 of them with the new
hermetically sealed package by Vandoren, ensuring a relative humidity
between 45 and $70\%$, according to the manufacturer), without moistening
the reed.

Each reed was subject to 2 series of measurements:

\begin{itemize}
\item \textbf{Series A} (asymmetrical excitation: see Fig.\ref{fig_fig2a}):
the right half of the mouthpiece chamber was filled with modeling clay to
ensure a good excitation of antisymmetrical modes. 184 holograms were made
ranging from 1.4 to 20 kHz (sinusoidal signal), by steps of 25 cents. The
amplitude of the excitation signal was exponentially increased in the
range 1.4 to 4 kHz, from 0.5 to 16 V, then kept constant at 16 V up to 20
kHz. This crescendo limits the amplitude of vibration of the first two
resonances of the reed. The temperature was not measured (about 20$^{\circ }$C).

\item \textbf{Series B }(symmetrical excitation: see Fig.\ref{fig_fig2b}):
the modeling clay was removed. The protocol is otherwise identical to this
of the first series. The reeds were inadvertently exposed during one night
to the very dry and warm air from the optical laboratory between the two
series of measurements. The reeds lost between 2 and $4\%$ of their mass. In
what follows we try to interpret the influence of this fact. The temperature was around 23-25$^{\circ }$C.
\end{itemize}

\subsubsection{ Nomenclature of normal modes}

\begin{figure}[tbp]
\begin{center}
\includegraphics[width=13cm,keepaspectratio=true]{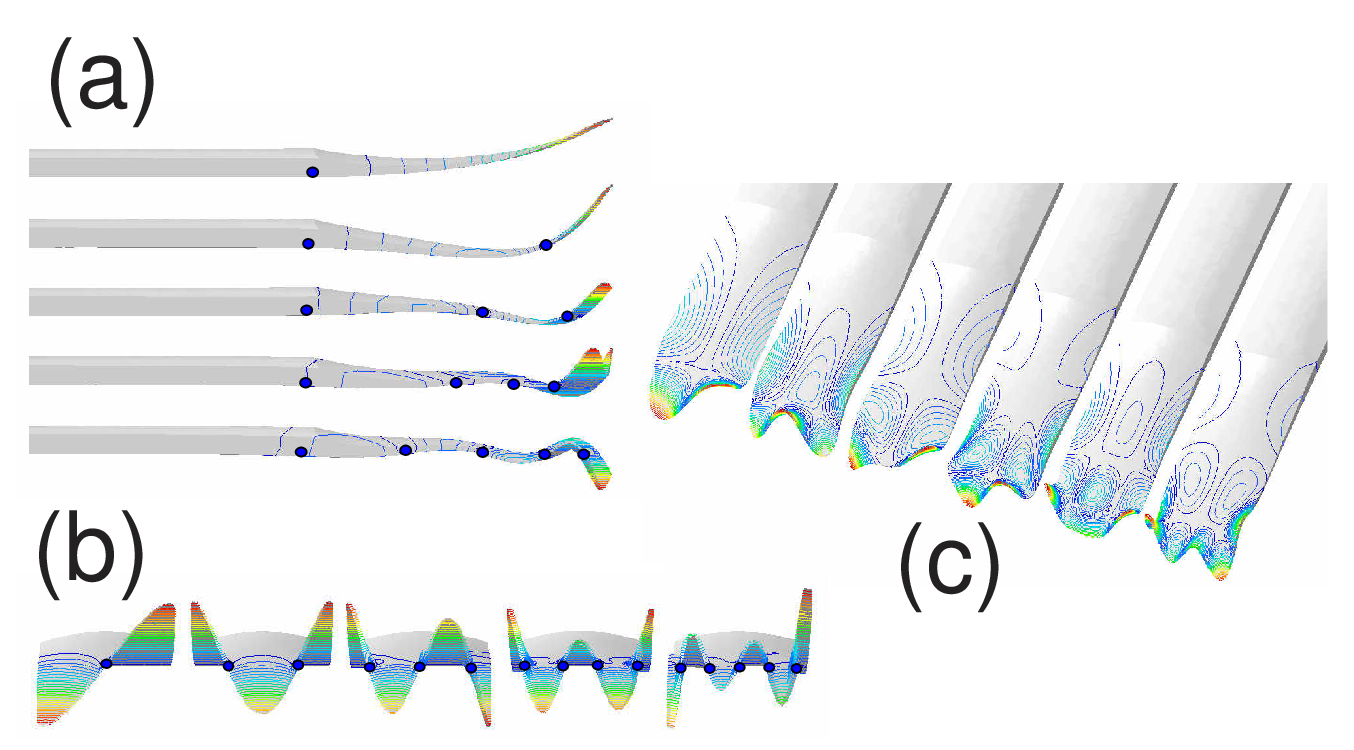} %
\end{center}
\caption{ (a) Flexural modes $F1$, $F2$, $F3$, $F4$ and $F5$. Side view. (b)
transversal modes $T1$, $T2$, $T3$, $T4$, $T5$. Front view. (c) Generic
modes: $X1$, $X2$, $X3$, $X4$, $X5$ and $X6$. View from above. The intersections of nodal lines with
the sides of the reed are symbolized by the blue dots.}
\label{fig_fig3abc}
\end{figure}

\begin{figure}[tbp]
\begin{center}
\includegraphics[width=10cm,keepaspectratio=true]{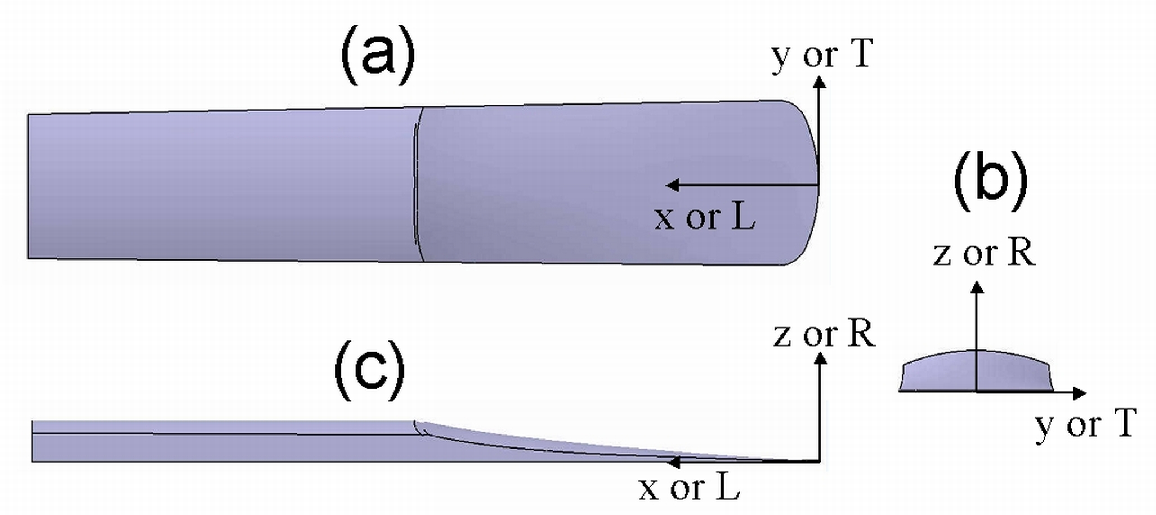}
\end{center}
\caption{ The clarinet reed: coordinates system. (a) top view, (b) front
view, (c) side view. $x$, $y$, $z$: Cartesian axes of the object. L, T, R: axes of the orthotropic material (L: longitudinal, T: tangential, R: radial). In the software Catia used for the simulations, the
orthotropic model is Cartesian and not cylindrical. Therefore  an exact
equivalence between $x,$ $y$, $z,$ and L, T, R respectively can be assumed
since we observed no important deviation between the direction of the grain and the axe of symmetry
of the reed. }
\label{Fig_fig_5}
\end{figure}

Distinguishing 3 morphological classes, we classify the modes of
a clarinet reed as follow : i) The
\textquotedblleft flexural\textquotedblright\ (or \textquotedblleft
bending\textquotedblright, or \textquotedblleft
longitudinal\textquotedblright ) modes, listed below $F$, whose frequencies
mainly depend on the longitudinal Young modulus ($E_{L}$) and polarized mainly in the $z$ axis, ii) the
\textquotedblleft transversal\textquotedblright\ (or \textquotedblleft\
torsional\textquotedblright , or \textquotedblleft
twisting\textquotedblright\ ) modes, listed below $T$, mainly dependent on
the shear modulus in the longitudinal / tangential plane ($G_{LT}$), and
iii) the \textquotedblleft generic\textquotedblright\ (or \textquotedblleft
mixed\textquotedblright ) modes, listed below $X$, sensitive to both moduli $%
E_{L}$ and $G_{LT}$ (see Fig.\ref{fig_fig3abc}). A subclass of flexural modes may be distinguished: the
\textquotedblleft lateral\textquotedblright\ modes (listed below $L$),
polarized mainly in the $y$ axis (see Fig. \ref{Fig_fig_5}). These modes were
not observed in our study.

The modes have been numbered after the order of
increasing frequencies
from a preliminary modal analysis we performed. In our analysis, however, the identification of a mode
is based upon morphological criteria. As a matter of
fact, the mode number and the order of
observed frequencies are not necessarily identical for all reeds.

Strictly speaking, the optical method only allows to observe the resonance
frequencies of the reed and not the eigenfrequencies. Therefore the observed
deformation patterns are a priori not identical to the eigenmodes of the
reed. Nevertheless in practice no major differences have be found between
the computed eigenmodes (see Section \ref{NuSi}) and the observed or computed
deformation for a forced asymmetrical excitation at the corresponding frequency. For this
reason we use the terminology \textquotedblleft mode\textquotedblright\ for
the maximum amplitude of the response of the reed to a forced excitation.
This is somewhat abusive, because the small shift between the resonance
frequencies due to damping and the eigenfrequencies computed by FEM, without
damping, is ignored. Besides damping, the acoustic load is also able to shift the resonance frequencies. We assume that this discrepancy is approximatively the same for all reed.

\subsubsection{Analysis of holograms; mode identification}

\begin{figure}[tbp]
\begin{center}
\includegraphics[width=14cm,keepaspectratio=true]{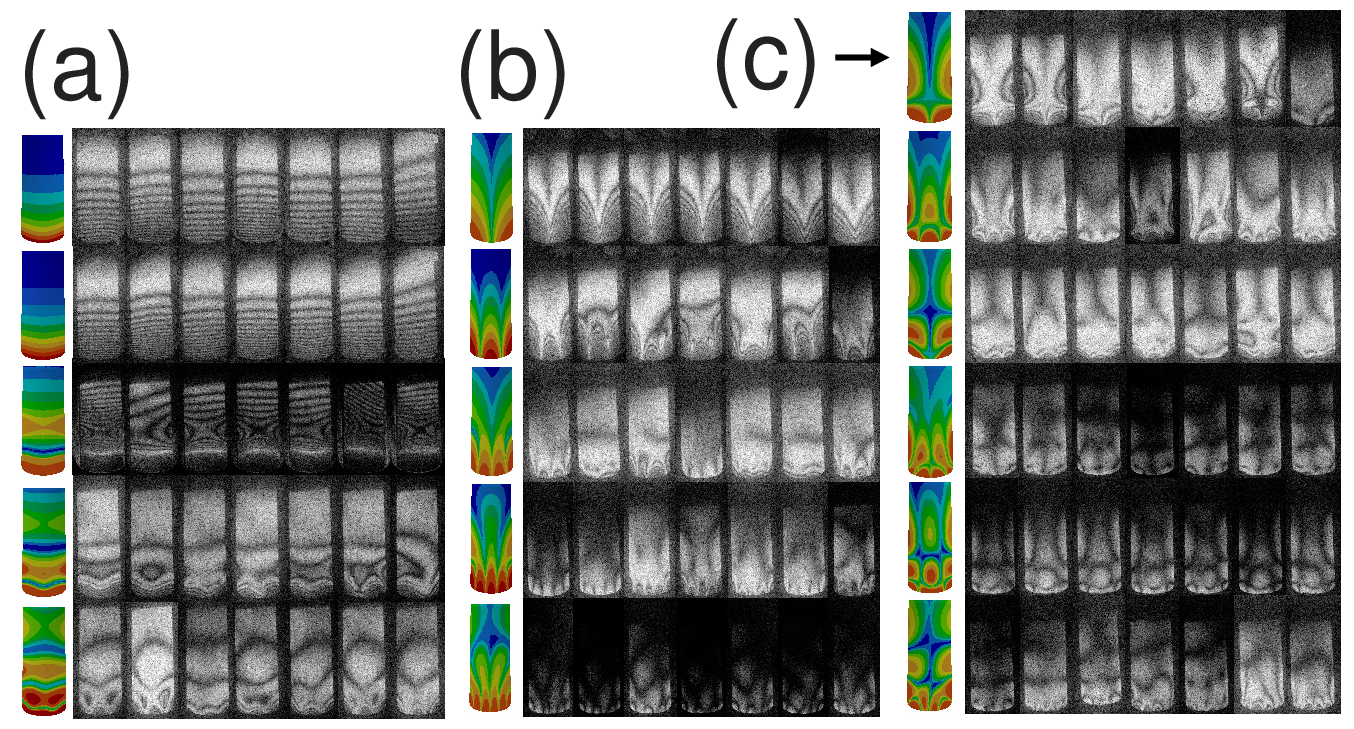}
\end{center}
\caption{ Qualitative comparison with FEM computation, and typical
variability of the experimental results.\ (a) First row: Quasi-static
pattern (at 605 Hz, with strong excitation by the LS). 2nd to 5th row: Flexural modes $F1$, $F2$, $F3$ and $F4$%
. Leftmost column: numerical simulation of eigenmodes by FEM. Columns 2 to
7: Arbitrary selection representing the observed variability (The first two
rows correspond to the same selection of reeds). Notice the marked
asymmetries and the differences in the curvature of the interference fringes
near the tip of the reed. (b) Rows 1 to 5: transversal modes $T1$, $T2$, $T3$%
, $T4$ and $T5$ (probably). Columns: see (a). (c) Rows 1 to 6: Generic modes
$X1$, $X2$, $X3$, $X4$, $X5$ and $X6$. The identification with $X6$ is
sometimes unlikely. Columns: see (a).}
\label{fig_fig4abc}
\end{figure}

More than 30000 holograms were made for this study and analyzed as follows: The picture where the
number of interference fringes is locally maximum is determined. For some cases, we
chose the hologram that is most similar to our numerical simulations
(by Finite Element Method, see Section \ref{NuSi}) or to other holograms
(see Fig.\ref{fig_fig4abc}). The holograms corresponding to an acoustical
resonance of the system, present at the same frequency (4309 Hz) for all reeds,
have been eliminated. The identification of the different patterns to those
calculated by FEM was often quite simple. An exception have been encountered
for $F3$ and $T3$, whose frequencies were often so close that our
identification is sometimes uncertain. More sophisticated techniques would
certainly solve this problem. Notice that other boundary conditions (e.g.
with clamping closer to the tip of the reed) would also easily separate
these two modes. The frequency of some higher modes could not always be
measured, either because their frequency was beyond 20 kHz, or because their
pattern could not be clearly identified.

\section{Statistical analysis of resonance frequencies\label{ModalFreq}}

4 flexural, 5 transversal, and 6 generic modes have been identified, namely all 15
first modes of the reed, excluding lateral modes. This number is
significant, compared to the 4 modes detected by Pinard \emph{et al.} \cite%
{pinard2003musical}. The 6th mode ($L1$) could not be identified, as it is a
lateral mode (flexural mode moving mainly in the $y$ axis), not excited by our loudspeaker. We tried to observe it by
rotating the mouthpiece to the side, without success.
Notice that higher modes could probably
be identified using an ultrasonic loudspeaker.
\subsection{Statistics}
The statistics are displayed on Fig. \ref{modes2} and detailed in Appendix \ref{statistic}, with the analysis of correlations.  For 14 measurements of resonance frequencies of the two series, identification of the
total number of reeds (55) have been done. For other measurements,
identification has been done only for a part of this number. The
value of the ratio of the standard deviation $\sigma $ to the mean
value $\mu $, i.e. the relative standard deviation, is found to be
between 2 and 5\% (about $1/3$ tone). If we admit Gaussian
distribution for the measured frequencies, $99\%$ of the observations
typically range about $\pm 1$ tone ($\pm 200$ cents) around the mean value (i.e. $\mu \pm 3\sigma $%
), for all frequencies.

The identification of the mode $X6$ is uncertain: it seems to appear for
frequencies lower than those of our simulations. Mode $T5$ is on the limit
of the range we studied: this explains the small value of the standard
deviation.

Between series A and B, the flexural modes $F1$ to $F4$ lower their
mid range, while the transversal modes slightly increase it. The difference
between the two series probably lies mainly in the drying of the reeds, and
this seems to have a statistically significant effect. This is surprising,
because drying decreases the density of the reed, and theoretically this
should proportionally increase all frequencies. In addition, according to
Obataya \emph{et al.} \cite{obataya1999c_acoustic}, drying is expected to
increase $E_{L}^{\prime }$ (at least around 400 Hz), which should also
increase the resonance frequencies. However Chevaux \cite{Chevaux_1997} observed
that drying diminishes $E_{L}^{\prime }$ for material extracted from the
inner side of the cane and augments slightly $E_{L}^{\prime }$ for material
extracted nearer from the outer side (for cane suitable for oboe reeds), at
least in the frequency range 100-500 Hz.

The hypothesis of an influence of the excitation method on the resonance
frequencies seems unlikely, as well as the hypotheses of a poor
reproducibility of the position of the reed on the mouthpiece between
measurements or of the modification of the acoustic load, due to the modeling clay.

\begin{figure}[tbp]
\begin{center}
\includegraphics[width=7cm]{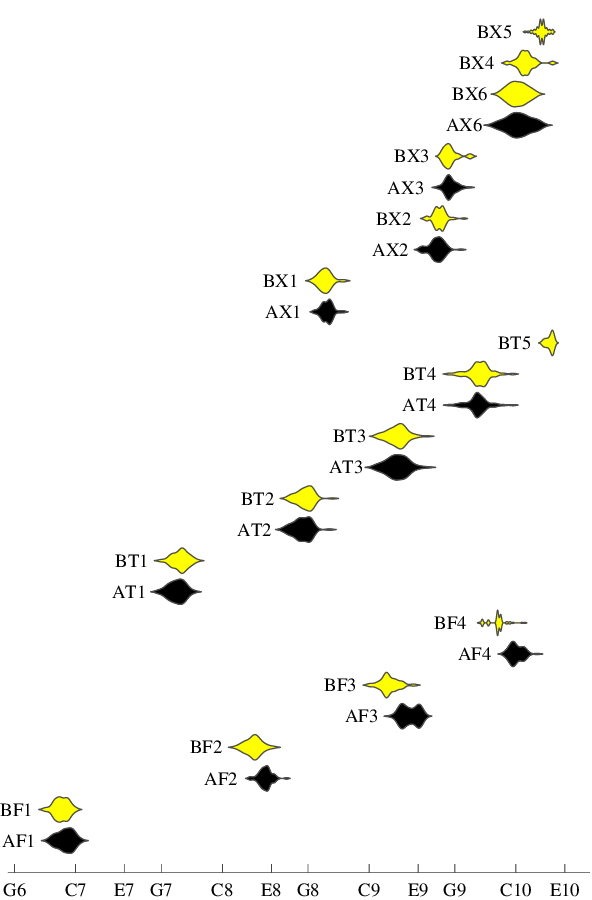}
\caption{Observed resonance frequencies for the different reeds of each series, according to the results of Table \ref{Table_table_1}. AF1 signifies mode F1, series A. The series A is in black, and the series B in white (yellow online). In abscissa the frequencies in logarithmic (musical) scale. From the left to the right, the different notes correspond to the following frequencies: 1568, 2093, 2637, 3136, 4186, 5274,
6272, 8372, 10548, 12544, 16744, 21096 Hz}
\label{modes2}
\end{center}
\end{figure}

\subsection{Principal component analysis}

\label{pcaDefinition}

Principal Component Analysis (PCA) is mathematically defined as an
orthogonal linear transformation transforming the data to a new coordinate
system, such that the greatest variance by any projection of the data comes
to lie on the first coordinate (called the first principal component or
first factor), the second greatest variance on the second coordinate, and so
on \cite{WikiPCA}.
Theoretically PCA is the optimum linear transform for given data in terms of
least squares. PCA is based upon the calculation of the eigenvalue
decomposition of the covariance (or of the correlation) matrix (see e.g. \cite{jolliffe2002principal}%
).

A PCA has been performed using the FACTOR module of SYSTAT \cite{Systat}. The 14 variables
(observed frequencies) presenting complete measurements for all reeds have
been selected (all variables having 55 identified pattern, $N_{A}$ or $N_{B}=55$,
see Table \ref{Table_table_1}). Frequencies are rated in cents.

The 4 largest eigenvalues have been selected. They capture $91.2\%$ of the
total variance of our sample (respectively $53.6\%$, $21.4\%$, $10.8\%$ and $%
5.4\%$ for each factor). A fifth factor would capture only $2.5\%$ variance
more. The 14-dimensional data have been linearly projected onto a
4-dimensional factor space.

The factor space can afterwards be orthogonally rotated, for instance for
maximizing the correlations between rotated factors and observed variables.
In the studied case, no a priori knowledge about the orientation of the
factor space is available. For an easy comparison, using the VARIMAX algorithm, we choose to maximize the
correlations between rotated factors and all available variables (observed resonance frequencies and
theoretical components from the model described hereafter in Section \ref{modH4}).

\begin{table}[tbp]
\centering%
\begin{tabular}{|c|c|c|c|c|}
\hline
& $factor_1$ & $factor_2$ & $factor_3$ & $factor_4$ \\
\hline
AT2 & \textbf{0.973} & 0.054 & 0.078 & -0.008 \\
BT2 & \textbf{0.953} & 0.055 & -0.025 & 0.085 \\
AT3 & \textbf{0.898} & 0.198 & -0.237 & -0.032 \\
AX2 & \textbf{0.853} & 0.273 & -0.073 & -0.123 \\
AT1 & \textbf{0.776} & 0.017 & 0.577 & 0.087 \\
BT1 & \textbf{0.762} & -0.033 & 0.541 & 0.177 \\
AX3 & \textbf{0.761} & 0.472 & 0.090 & 0.217 \\
AX1 & \textbf{0.740} & 0.451 & 0.356 & 0.115 \\
AF3 & 0.223 & \textbf{0.891} & 0.076 & 0.080 \\
AF2 & 0.143 & \textbf{0.791} & 0.519 & 0.064 \\
AF1 & 0.085 & 0.360 & \textbf{0.870} & -0.141 \\
BF1 & 0.104 & 0.385 & \textbf{0.835} & 0.209 \\
BF3 & 0.059 & 0.596 & 0.050 & \textbf{0.755} \\
BF2 & 0.147 & 0.561 & 0.290 & \textbf{0.710} \\ \hline
$e_{1}$ & \textbf{0.958} & -0.080 & 0.105 & 0.005 \\
$e_{2}$ & 0.059 & \textbf{0.969} & 0.082 & 0.098 \\
$e_{3}$ & -0.075 & -0.073 & \textbf{0.971} & 0.042 \\
$e_{4}$ & -0.009 & -0.081 & -0.039 & \textbf{0.979} \\ \hline
\end{tabular}%
\caption{ Correlation (loadings) between rotated factors from PCA and : i)
variables (measured resonance frequencies) or ii) components from viscoelastic
model ($\boldsymbol{e}[n]$,  see Section \protect\ref{modH4}), for comparison, sorted in reverse order of
magnitude. In \textbf{bold:} \ greater correlation for each variable. }
\label{Table_table_2}
\end{table}

We performed also a PCA separately for each measurement series (A and B: 9
and 5 variables, respectively). From series A we detected 3 important
factors capturing $90.8\%$ of the variance (56.9, 23.0, $11.0\%,$
respectively). From series B we detected also 3 important factors capturing $%
94.1\%$ of the variance (54.0, 26.8, $13.3\%$, respectively). A 4th factor
would capture only $3.6\%$ more for series A and $3.4\%$ for series B. One
factor seemingly disappeared, compared with the PCA performed on both
series. A hypothesis is that this factor is related to the hygrometric
change between the two series.

>From Table \ref{Table_table_2}, we see that all transversal and generic
modes are well correlated with $factor_1$; $factor_2$ correlates with high
frequency flexural modes of both series (however notably better with those
of series A); $factor_3$ well correlates with $F1$ of both series (and
somewhat with other low frequency modes: AT1, BT1 and AF2), whereas $factor_4$
correlates quite well with high frequency flexural modes of series B.

\subsection{Conclusions from the statistical analysis}\label{ConclStat}

    Fifteen modes of vibration of the clarinet reed have been
observed, while previous studies
investigated 4 to 5 modes only \cite{Guimezanes_2007, pinard2003musical}. The observed resonance frequencies are often highly correlated, especially those among the
\textquotedblleft transversal\textquotedblright\ modes and, to a lesser extent, those among the \textquotedblleft flexural\textquotedblright\  modes.
The nominal reed strength is surprisingly better correlated with the
frequencies of \textquotedblleft transversal\textquotedblright\ modes as with those of \textquotedblleft flexural\textquotedblright\  modes. The flexural modes within the same series are poorly correlated.

A principal component analysis of the resonance frequencies identifies 4 main
factors, capturing $91.2\%$ of the variance of the sample. The data can therefore be reconstructed with 4 uncorrelated factors only (error: $RMSD= 21.8$ cents, see Appendix \ref{MSD} and \ref{reconstr}). The effect of hygrometric change between both measurement series can seemingly be described with 1 factor only.

These statistical facts offer a guidance for modeling appropriately the mechanics of the clarinet reed.

\section{Development of a mechanical model\label{NuSi}}

\subsection{Choice of a viscoelastic model}

In the present study our concern is to develop a model with a minimal number of physically related components, that adequately reconstructs the observed resonance frequencies of our reeds. We presume that these components offer an objective characterization of the material composing each reed. A sensitivity analysis by FEM calculation
assuming an orthotropic, elastic material has been conducted, and showed that
the longitudinal Young modulus $E_{L}$ and the longitudinal / transverse
shear modulus $G_{LT}$ play a leading role. Nevertheless taking into account the previous result of 4 factors given by PCA, we will see that an elastic model is not sufficient to establish a satisfactory model with 2 degrees of freedom only (i.e. variables $E_{L}$ and $G_{LT}$ per series.  Therefore a viscoelastic model is sought.

It is well known that the stiffness of natural materials like wood or cane varies with the frequency
of the applied stress and with the temperature. The material is stiffer at
low temperature and at high frequency. At low frequency or high temperature
the material is almost perfectly elastic and reaches the rubbery modulus. At
high frequency or at low temperature the glassy modulus is reached; the
material is almost perfectly elastic, also, but stiffer. At mid frequency or
mid temperature, the apparent modulus (called storage modulus, i.e. the real
part of the complex Young modulus for this frequency) is between the two
values. For a particular frequency, called relaxation frequency, the storage
modulus is exactly at the average of glassy and rubbery moduli. Around this
frequency dissipation is maximum. Once the characteristic curve is known
(for given temperature and different frequencies, or for given frequency and
different temperatures), the Arrhenius equation\footnote{%
The shift in relaxation time is: $Ln(shift)=\frac{E_{a}}{R}\left(
\frac{1}{T}-\frac{1}{T_{ref}}\right) $ , where $E_{a}$ is the activation
energy, $R$ is the gas constant (8.314 J/K~mol) and $T$ and $T_{ref}$ the
absolute temperatures in K. For instance, a shift of +10$^{\circ }$C from a
reference temperature of 20$^{\circ }$C decreases the relaxation time by 16$%
\%$, if $E_{a}$ is 13~kJ/mol.} offers usually an adequate estimate of the
stiffness for any frequency and any temperature, within a quite broad range
\cite{roylance2001engineering}.

The determination of the mechanical parameters of a natural material
 requires to determine for each axis of the
orthotropic material the
value of 3 parameters (Young modulus, shear modulus and Poisson's ratio).
These 9 parameters may exhibit viscoelastic behavior, requiring
theoretically for each one the fit of a viscoelastic model, such as the
general linear solid (also called Zener model or 3-parameter model, see \cite%
{roylance2001engineering, chaigne2008acoustique,
dill2006continuum,gaul2007experimental})\footnote{ Other multidimensional viscoelastic models could be also considered. In order to
fit a wide range of frequencies (more than 2 decades), a 4-parameter model
with fractional derivative would be required \cite{gaul2007experimental}. In addition, these
parameters are known to be sensitive to moisture content. Moreover, the cane is not homogeneous. The stiffness varies in radial direction \cite%
{Chevaux_1997} and local irregularities may be important, as shown by J.-M.
Heinrich \cite{Heinrich_1991}.}.
 The chosen model is based on 6 parameters (3 parameters for both variables $E_{L}$ and $G_{LT}$). Therefore the viscoelastic model (Section \ref{ViMo}) has many degrees of freedom (6, for each of the 2 series of measurements, compared to the 4 factors detected by PCA for both series), for solving adequately the inverse problem (see Section \ref{invProb} for the reduction of the number from 12 to 4).

\subsection{Computation method}

Considering viscoelasticity leads to complex modes with complex
eigenfrequencies. Compared to the non-dissipative, elastic case computed by
FEM, the main consequence of viscoelasticity, besides dissipation, is that
stress and strain are not in phase. For sake of simplicity, we limit the
computation to eigenfrequencies only, and assume that they depend on the
storage moduli only (i.e. dissipation has a negligible influence). Having
reduced the viscoelastic problem to an associated elastic one, the elastic
solution may be used (see e.g. Ref. \cite{roylance2001engineering}). In
order to compute the resonance frequency $\omega _{r}$ after an elastic model,
according to Ref. \cite{gaul2007experimental}, we admit that $E\simeq
E^{\prime }(\omega _{r})$, where $E^{\prime }(\omega _{r})$ is the real part of
the complex modulus in the frequency domain. This hypothesis implies that
the calculation of the eigenfrequencies from the values of the storage
modulus is done by an iteration procedure, and allows to use a FEM software (Catia) which does not allow  computing with frequency-dependent coefficients.

Therefore we first present results of FEM simulations (Section \ref{section_III_A}), assuming an elastic and orthotropic behavior of the reed (modeled in Section \ref{reedShape}). This helps to identify the
modes in experiments, and allows  obtaining a fit formula (Section \ref{algorithme}) for computing the 11 lower resonance frequencies with respect to two parameters only, $E_{L}$ and $G_{LT}$, detected after a sensitivity analysis (Section \ref{sensiAna}). The fit formula, called "metamodel", is then used in the iterative procedure for the computation of the viscoelastic model. It allows a great reduction of computation time, compared to the FEM, and this is very useful for the inverse problem. Such a metamodel could be directly computed for the viscoelastic model with an appropriate software, but starting with the elastic model simplifies the fitting procedure.

\subsection{Elastic model}

\label{section_III_A}

\subsubsection{Modeling the reed}\label{reedShape}

The clarinet reed is defined in a Cartesian axis system $x,y,z$ (see Fig. %
\ref{Fig_fig_5}). The origin is located in the bottom plane, at the tip of
the reed. The material is defined as 3D orthotropic and assumed to be
homogeneous, whose longitudinal direction L is parallel to the $x$ axis, the
tangential direction T parallel to the $y$ axis and the radial direction R
parallel to the $z$ axis\footnote{Do not confuse the morphological mode classes $L1$, $L2$, $T1$, $T2$, $T3$ and $T4$ with the axes L and T of the orthotropic material. }.

The dimensions in the $xy$ plane are consistent with the measurements given
by Facchinetti \emph{et al.} \cite{facchinetti2003numerical}. The heel of
the reed is made out of a cylinder section, diameter 34.8~mm, maximum
thickness 3.3~mm. The shape of the reed is defined in Appendix \ref%
{shape}.

During playing, the reed has two contact surfaces with the ligature. For the
present simulations, the reed is clamped in the same way than for normal
playing, on two rectangular surfaces $23\times 1$~mm, spaced laterally by
5~mm, 38.2~mm from the tip of the reed, simulating the contact surfaces on
the Vandoren Optimum ligature. However, unlike normal playing, the whole vamp of the reed is free to vibrate (see Figure \ref{fig_setup}).

For the simulations, the \textquotedblleft Generative Part Structural Analysis\textquotedblright\ module by
Catia v.5.17 (Dassault Technologies) is used, with mesh Octree3D, size 2~mm,
absolute sag 0.1~mm, parabolic tetrahedrons. The generated mesh involves
5927 points, allowing both a good accuracy and a reasonable computing time
(around 35 seconds).

\subsubsection{Sensitivity analysis of elastic coefficients}\label{sensiAna}

\begin{table}[tbp]
\centering
\begin{tabular}{|c|c|c|c|c|c|}
\hline
Coefficient & $F$ & $T$ & $X$ & $L$ & All modes \\ \hline
$E_L$ & \textbf{0.4087} & 0.1053 & 0.1835 & 0.2093 & 0.2235   \\
$G_{LT} $ & 0.0140 & \textbf{0.2681} & 0.1962 & 0.1067 & 0.1575  \\
$E_T$ & 0.0076 & \textbf{0.0976} & 0.0741 & 0.0166 & 0.0562   \\
$G_{LR} $ & 0.0438 & 0.0131 & 0.0257 & \textbf{0.0818} & 0.0341   \\
$E_R$ & 0.0176 & 0.0120 & 0.0135 & \textbf{0.0662} & 0.0207   \\
$G_{TR}$ & 0.0046 & 0.0092 & 0.0077 & \textbf{0.0215} & 0.0091   \\
$\nu_{TR}$ & 0.0015 & 0.0009 & 0.0015 & \textbf{0.0054} & 0.0019   \\
$\nu_{LT}$ & 0.0018 & \textbf{-0.0031} & 0.0004 & 0.0007 & -0.0001  \\
$\nu_{LR}$ & 0.0009 & 0.0002 & 0.0005 & \textbf{0.0013} & 0.0007   \\ \hline
\end{tabular}%
\caption{ One-At-a-Time sensitivity study by FEM: Averaged ratio between relative change
in frequency and relative change for each elastic coefficient (i.e. $\pm
10\% $), sorted by decreasing order of magnitude, for the first 16
eigenmodes. $F$: flexural modes ($F1$ to $F4$), $T$: transversal modes ($T1$ to $T4$),
$X$: generic modes ($X1$ to $X6$), $L$: lateral modes ($L1$ and $L2$; these modes were
not observed in our study), All modes: averaged ratio over all modes. In
\textbf{bold}: maximum absolute value for each coefficient of the
orthotropic material: $E_{L}$, $E_{T}$ and $E_{R}$: Young moduli; $\protect%
\nu _{LT}$, $\protect\nu _{LR}$ and $\protect\nu _{TR}$: Poisson
coefficients; $G_{LT}$, $G_{LR}$ and $G_{TR}$: shear moduli. }
\label{Table_table_3}
\end{table}

 For selecting the most relevant
parameters, we conducted a One-At-a-Time sensitivity analysis \cite{saltelli2008global}, varying each coefficient by
$\pm 10\%$ and computing the first 16 modes, based on the following
reference values: $E_{L}=$14000~MPa, $E_{T}=E_{R}=$ 480~MPa, $\nu _{LT}$ $%
=\nu _{LR}=\nu _{TR}=$0.22, $G_{LT} =$ 1100~MPa, $G_{LR}=G_{TR}=$ 1200~MPa.
The density $\rho $ was set to 520 kg/m$^{3}$, according to the estimation
by Guimezanes \cite{Guimezanes_2007}. The results are shown in Table \ref%
{Table_table_3}. Notice that $E_{L}$ and $G_{LT}$ plays a decisive role,
while $E_{T}$ plays a marginal role and all other parameters have an almost
negligible influence on the resonance frequencies. As a consequence, the moduli $%
E_{L}$ and $G_{LT}$ are the variables retained in the model. The approximate
value of $E_{T}$ has been estimated according to the morphology of the
patterns of higher order modes. This value is consistent with measurements
given by Spatz \emph{et al.} \cite{spatz1997biomechanics}.

Notice that these results show the validity of a 2D approach, the reed being
modeled as a thin plate. This should be used for further studies.

\subsubsection{Metamodel approximating the resonance frequencies}

\label{algorithme}
The following analytic formula ("metamodel") predicts quickly and efficiently the resonance frequencies of a clamped/free clarinet reed. It was established in the following way: Frequencies of the first 16 modes were
computed by FEM, according to a network of 92 separate pairs of values for $%
E_{L}$ and $G_{LT}$, ranging from 8000 to 17000~MPa and 800 to 1700~MPa,
respectively. The other elastic coefficients were held constant, according
to the reference values cited above. For the range of simulation values, this arbitrary formula (developed by trial and error) provides a very good fit (generally better than $\pm5$ cents, see Table \ref{tab9}). Expected resonance
frequencies $f$ are first found in $cents$ ($FC$) from the note F6 (1396.9~Hz), and finally in $Hz$:
\begin{eqnarray}
f(m,E_{L},G_{LT}) &=&1396.9\times 2^{FC/1200}\text{, where}  \label{eq1} \\
FC
&=&a_{m,0}+a_{m,1}\,E_{p}+a_{m,2}G_{p}+a_{m,3}\,E_{p}\,G_{p}+a_{m,4}%
\,E_{p}^{2}+a_{m,5}G_{p}^{2}\text{,}  \notag \\
E_{p} &=&E_{L}~^{-0.66643}\text{ and\ }G_{p}=G_{LT}~^{0.7627}.  \notag
\end{eqnarray}%
The index $m$ is the number of the mode defined in Appendix \ref{meta}, Table \ref%
{tab9}, where the values of the coefficients $a_{m,q}$ are given ($E_{L}$
and $G_{LT}$ are expressed in MPa).

The influence of the density is easy to predict:
frequencies vary proportionally to $\rho ^{-1/2}$. The computing cost of this metamodel is about $10^7$ times lower than with FEM, largely simplifying the inverse problem.

\subsubsection{Efficiency of the metamodel}
Equation (\ref{eq1}) can be used to estimate the values of $E_{L}$ and $G_{LT}$,
providing a faithful reconstruction of the observed resonance frequencies.
Theoretically these values could be computed for any pair of modes, after
their respective observed frequencies. Unfortunately, this method gives no consistent results. A least squares fit is a more robust technique for such a computation. This leads however to systematic errors in the predicted frequencies: low-order modes are systematically overestimated, while high-order modes are underestimated. This can be corrected by adjusting the coefficients $a_{m,0}$ (from Table \ref{tab9}), but this cannot explain the bad correlation among flexural modes within the same series (see Table \ref{TableCorr}). According to the elastic model, these correlations should be in all cases greater than 0.998. A hypothesis for resolving this contradiction is that the moduli are varying with the frequency in an individual way for each reed. Thus in the next paragraph we consider  a viscoelastic model, where $E_{L}$ and $G_{LT}$ are frequency dependent. This leads to the addition of some
parameters, which are to our mind more important that the other elastic
coefficients. The
fit of such a model requires many observations at different frequencies, in
order to reduce the influence of measurements errors and of local
irregularities in the structure of cane.

Alternative hypotheses could be considered in this context, as damping, acoustic load \cite{facchinetti2003numerical}, local variations in stiffness or in density,  local deviations in thickness, compared to the assumed theoretical model. However, these hypotheses are probably unable to explain the hygrometric-induced individual variations we observed for each reed, thus our preference for the viscoelastic hypothesis.

\subsection{Viscoelastic model \label{ViMo}}

\begin{figure}[tbp]
\begin{center}
\includegraphics[width=4cm,keepaspectratio=true]{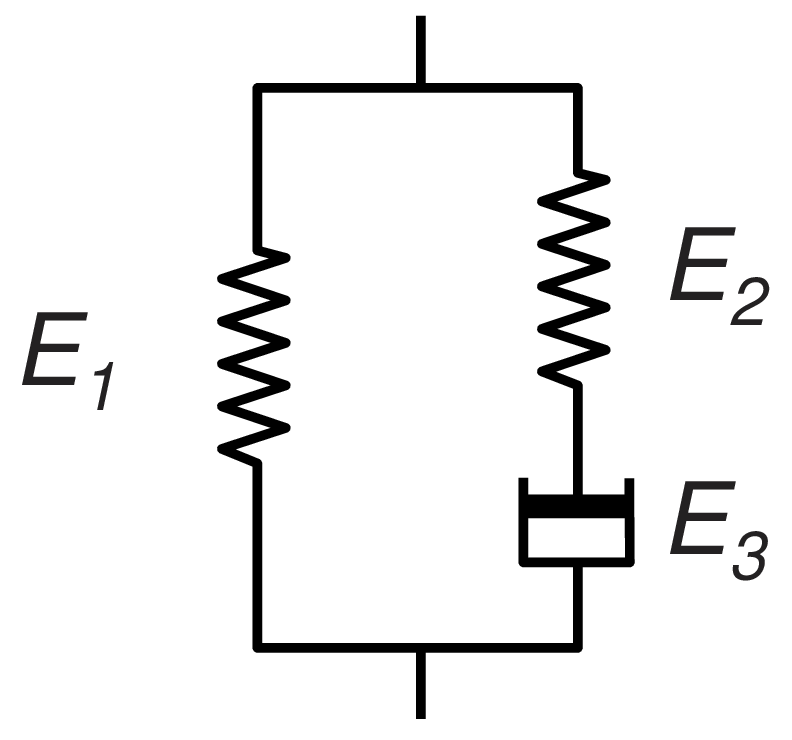}
\end{center}
\caption{ Schematic representation of the standard linear solid: two springs
$E_{1}$, $E_{2}$ and a dashpot $E_{3}$.}
\label{Fig_fig_6}
\end{figure}

In this section, a Zener model is considered (see e.g.
Refs \cite{roylance2001engineering, chaigne2008acoustique,
dill2006continuum, gaul2007experimental}). This model is applied to both moduli $E_{L}$ and $G_{LT}.$ The scheme of the standard
viscoelastic solid is presented on Fig.\ref{Fig_fig_6}, with two springs $%
E_{1}$ and $E_{2}$ and a dashpot $E_{3}$\footnote{This notation allows to write the parameters of the model as a vector, as required by the computations, but unfortunately it hides the fact that the nature of $E_3$ (dashpot) is physically different from $E_1$ and $E_2$ (springs).}. At low frequencies, $%
E_{2}$ and $E_{3}$ have practically no effect (the rubbery
modulus $E_{1}$ dominates). At high frequencies, $E_{3}$ has
practically no effect (the glassy modulus $E_{1}+E_{2}$ dominates). In the
frequency range near $E_{2}/(2\protect\pi E_{3})$ the
dissipation due to $E_{3}$ is maximal and the apparent modulus
(storage modulus) is in the mid-range. The stress $\sigma $ and the
strain $\varepsilon $ are related by the constitutive equation:

\begin{equation}
\sigma +\tau _{1}\dot{\sigma}=E_{1}(\varepsilon +\tau _{2}\dot{\varepsilon})
\label{eq2}
\end{equation}%
in which $\tau _{1}={E}_{3}/{E_{2}}$ is called the relaxation time and $\tau
_{2}={E}_{3}{(E_{1}+E_{2})}/{(E_{1}E_{2})}$ the retardation time. $E_{1}$ is
called rubbery modulus and $E_{1}+E_{2}$ glassy modulus. In harmonic regime,
for an angular frequency $\omega $, the Young modulus is complex:
\begin{equation}
E^{\ast }(\omega )=E_{1}+E_{2}-\frac{E_{2}^{2}}{E_{2}+j\omega E_{3}}%
=E_{1}+E_{2}+\frac{-E_{2}+j\omega E_{3}}{1+\left( \omega E_{3}/E_{2}\right)
^{2}}  \label{eq3}
\end{equation}%
The second formulation separates the real part ($E^{\prime }(\omega )$:
storage modulus) and the imaginary part ($E^{\prime \prime }(\omega )$: loss
modulus) of $E^{\ast }(\omega )$. The storage modulus can thus be written
as:
\begin{equation}
E^{\prime }(\omega )=E_{1}+E_{2}-\frac{E_{2}^{3}}{E_{2}^{2}+\omega
^{2}E_{3}^{2}}=E_{1}\frac{1+\omega ^{2}\tau _{1}\tau _{2}}{1+\omega ^{2}\tau
_{1}^{2}}.  \label{eq4}
\end{equation}%
Notice the properties:
\begin{eqnarray}
&&E^{\prime }(0)=E_{1}\text{ ; \ }E^{\prime }({1}/{\tau _{1}})=E_{1}+E_{2}/2
\notag \\
&&E^{\prime }(\infty )=E_{1}+E_{2}\text{ ; }\frac{\partial E^{\prime }}{%
\partial \omega }\Big(\frac{1}{\tau _{1}}\Big)=\frac{E_{3}}{2}.  \notag
\end{eqnarray}%

 For the sake of
simplicity, the parameters corresponding to $E_{L}$ and $G_{LT}$ are denoted $E_{1},$ $E_{2}$, $E_{3}$, and $G_{1}$, $G_{2}$, $G_{3}$, and the storage moduli given by Equation (\ref{eq4}) $%
E^{\prime }(\omega )$ and $G^{\prime }(\omega )$, respectively$.$ Therefore
for each reed (and each series), the model requires 6 parameters instead
of 2 (while experiments gave 4 main factors only for the whole set of
results).

>From the knowledge of the 6 parameters, the resonance frequencies can be
deduced by an iteration procedure. For each mode the starting point of the
iteration is the mean value $f^{(0)\text{ }}$of the experimental resonance
frequency (see Table \ref{Table_table_1}), then the storage moduli are
deduced from Equation (\ref{eq4}), then a new value $f^{(1)}$ by using Equation (\ref%
{eq1}), etc... The convergence of the iteration method is fast, actually one
iteration is enough. This can be understood by the fact that the derivative
of the iterated function is small (notice that the two first rows of Table %
\ref{Table_table_3} correspond to the derivative of $E_{L}$ and $G_{LT}$ with
respect to frequency). If we give an arbitrary value, for instance $%
f^{(0)}=6000$, one iteration more is required for a comparable precision.
In all hypotheses (see Appendix \ref{Hypot}), we used one iteration only. This procedure allows the determination
of $f(m)$ from the coefficients $E_{1},$ $E_{2}$, $E_{3}$, $G_{1}$, $G_{2}$,
$G_{3}$ for a given reed and a given series.

\section{Inverse problem and selection of a robust model}\label{invProb}
\subsection{Simplification of the model by multiple regression}
In order to solve the inverse problem for each reed (and each series), we
use a classical Mean Squared Deviation method, from the experimental values
of the 11 resonance frequencies listed in Table \ref{tab9} (see Appendix \ref{meta}). The results  are given and discussed in Section \ref{results}.

However for each reed the viscoelastic model provides 12 parameters, i.e., 12 DoF, and  the PCA showed that this number needs to be reduced. Actually  this  model  conducts sometimes to non-physical results (negative rubbery modulus, for instance). This problem comes out because the observed resonance
frequencies are far from 0, so the rubbery modulus $E_{1}$ cannot be
estimated precisely. For that purpose   multiple regression (see Appendix \ref{DVM} for details) is used, by introducing correlations among parameters and reducing the degrees of freedom to a number of~4, called ``components", which are linearly related to the parameters. The 4 components are very similar to the 4 factors computed by the PCA, but, because Eqs. (\ref{eq1} and \ref{eq4}) are nonlinear, a small deviation is inevitable for optimal results. Factors and components are consequently strongly correlated (>0.95, see Table \ref{Table_table_2}).

We tested different hypotheses to establish a satisfactory robust model (denoted H1 to H9 and described in Appendix \ref{DVM}), together with the detailed computation method.  Each hypothesis leads to a given number of components related to the parameters through the regression coefficients.
Notice that the parameters and components depend on both the reed and series, while the regression coefficients which correlate the parameters are independent of the reed and the series.
A constant value elastic model (the parameters are fixed, and do not depend on the reed or the series, Hypothesis H1)  is not sufficient. Similarly  a 2-parameter elastic model ($E_{L}$ and $G_{LT}$, independent of the series, Hypothesis H2) is not sufficient as well.  Conversely a linear model without constraints (with 60 independent coefficients, corresponding to a matrix of order 12(4+1) elements\footnote{For this example, the component vector has 4 elements to be determined, plus a fifth element, which is a constant.}, Hypothesis H7) is not necessary, because many coefficients are very small. Eventually a 9 coefficient linear model (H4) has been found to be very satisfactory and is described  in Section \ref{modH4}. The following ideas were applied: after eliminating the very small coefficients and fitting the observations with the remaining coefficients, it is observed that the predictive quality of the model is almost not affected by the simplification. This was done step by step. The problem of non-physical values for certain parameters can be corrected  by setting that the damping parameters
$E_{3}$ and $G_{3}$ are constant, independent of the reeds and of the series
(typical values: $E_3 =$ 0.28 and $G_3 =$ 0.02). It
remains 4 parameters for each series and reed, $E_{1}$, $E_{2}$ and $G_{1}$%
, $G_{2}$ (i.e. 8 parameters for each reed). This ensures generally that these parameters fall in a
plausible range, when fitting the model.
Moreover a hierarchical structure can be introduced in the model, isolating the hygrometric component, bringing the remaining 3 components to a common basis (Section \ref{modH4}) and simplifying the problem (reduced to only 9  regression coefficients) and giving some insight in the data structure.

The $RMSD$ (Root Mean Square Deviation, see Appendix \ref{MSD}) is found to be
30.4 cents for Hypothesis H4, very close to 29.8 cents for H5 with 9 coefficients more.
Moreover the standard deviation of the residuals for Hypothesis H4 (and also Hypothesis H5)
varies very few over the different resonance frequencies (all around 30 cents).

The fit quality  cannot be considered as a perfect and definitive proof that our model
 reflects the true
values of the corresponding storage moduli. The influence of some missing parameters in the model should be examined (for instance differences in thickness between reeds,  non constant modulus $E_{T}$, non constant density $\rho$ or radial variation of $E_{L}$). Anyways,
the presented model reflects real mechanical differences between the reeds, very similar to those objectively detected by the PCA.

\subsection{Robust estimation of the parameters of the viscoelastic model (Hypothesis H4)}\label{modH4}

 Hypothesis H4 is chosen so that no coefficient can be removed without impacting notably the quality of fit. It can be thought as the minimal structure allowing an adequate reconstruction of the observed resonance frequencies, in conjunction with the viscoelastic model (Eq. (\ref{eq4})) and the metamodel (Eq. (\ref{eq1})). This minimal structure makes the model more robust against measurements errors, even if it probably introduces some bias.

As a first step, our concern is to eliminate the influence of the moisture content and to bring both series of measurements to a common basis (i.e. predict the effect of drying on the viscoelastic parameters of series B, the series A being taken as a reference). $e_1[n]$, $e_2[n]$, $e_3[n]$ and $e_4[n]$ are the 4 independent components characterizing the mechanical properties of the reed $n$. The choice of the notations is as follows: $e_k[n]$ is the $k$th element of the vector $\mathbf{e}[n].$, which depends on $n$.  These components are conditioned similarly to PCA as orthogonal factors: mean 0, standard deviation 1 and intercorrelation 0. The elimination of the moisture content can be achieved by reducing the components to a number of 3 for each series $s=1$ (series A) and $s=2$ (series B): for reasons explained in Appendix \ref{simplifiedModel}, these components are denoted  $\check{e}_1[s,n]$, $\check{e}_2[s,n]$ and $\check{e}_3[s,n]$ .

For series A, the components remain unmodified (series A is taken as reference):
\begin{eqnarray}
  \nonumber \check{e}_1[s=1,n]=e_1[n]\\
  \nonumber \check{e}_2[s=1,n]=e_2[n]\\
  \check{e}_3[s=1,n]=e_3[n]
\end{eqnarray}

For series B, the effect of drying on the components is predicted by:
\begin{eqnarray}
    \nonumber \check{e}_1[s=2,n]=c_{10}+e_{1}[n]\\
    \nonumber \check{e}_2[s=2,n]=c_{20}+\frac{1}{2}(e_{2}[n]+e_{4}[n])\\
    \check{e}_3[s=2,n]=c_{30}+e_{3}[n]
\end{eqnarray}

With this choice of components, the viscoelastic parameters of the model for series $s$ and reed $n$ can then be estimated as follows:

\begin{eqnarray}
       \nonumber E_{1}=d_{10} + d_{12}(\check{e}_2[s,n]+\check{e}_3[s,n])  \\
       \nonumber E_{2}=d_{20} + d_{12}(2 \check{e}_2[s,n]-\check{e}_3[s,n]) \\
       \nonumber E_{3}=d_{30} \\
       \nonumber G_{1}=d_{41}(6+\check{e}_1[s,n])  \\
       \nonumber G_{2}=2 d_{41}(3+\check{e}_1[s,n]-\check{e}_3[s,n]) \\
       G_{3}=d_{60}
               \label{EqH4}
\end{eqnarray}

This implies some other interesting relationships:
\begin{eqnarray}
       \nonumber E_{1}+E_{2}=d_{10}+d_{20} + 3 d_{12}\check{e}_2[s,n]  \\
       \nonumber G_{1}+G_{2}=d_{41}(12+3\check{e}_1[s,n]-2\check{e}_3[s,n]) \\
       \nonumber 2 E_{1}-E_{2}=2 d_{10}-d_{20} + 3 d_{12}\check{e}_3[s,n]  \\
                 2 G_{1}-G_{2}=2 d_{41}(3-\check{e}_3[s,n])
               \label{EqH4b}
\end{eqnarray}

Notice that the glassy modulus of $E_L$ (i.e., $E_{1}+E_{2}$) depends linearly only on $\check{e}_2[s,n]$. The quantities $2 E_{1}-E_{2}$ and $2 G_{1}-G_{2}$ depend linearly only on $\check{e}_3[s,n]$, however with opposed signs.

The values of the 9 coefficients are: $c_{10}$=1.011, $%
c_{20} $= -2.197, $c_{30}$=0.8294, $d_{10}$=10300, $d_{12}$=640.5, $d_{20}$%
=7309, $d_{30}$=0.2822, $d_{41}$=115.7, $d_{60}$=0.02038. The coefficients in Eq. (\ref{eq1}) are adjusted (in order to remove systematic errors) by adding to $a_{m,0}$ (from Table \ref{tab9}) the following values, for $m=1$ to 11:  -26.27, 32.24, -50.05, 4.80, -26.87, -65.28, -48.64, 0.07, -52.98, -76.61 and -111.55 cents.

The change in density between the two series of measurements was not measured precisely (about -2 to $-4\%$). In the model, density is considered as constant.
\begin{figure}[tbp]
\begin{center}
\includegraphics[width=9cm,keepaspectratio=true]{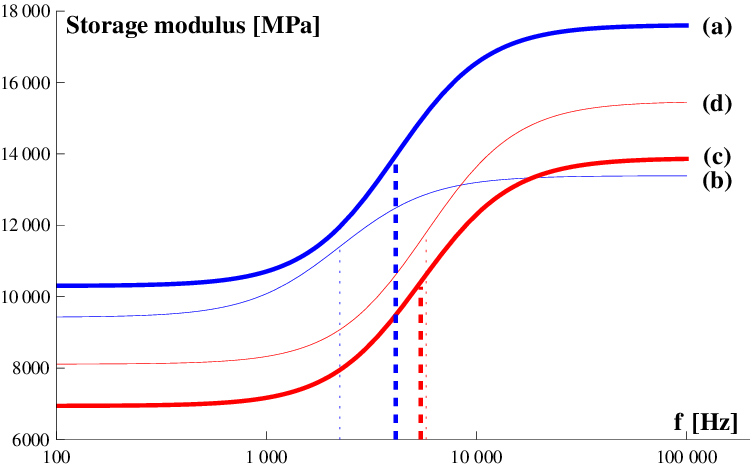} %
\end{center}
\caption{Hypothesis H4: plot of the storage moduli $E_{L}^{\prime }(\protect\omega )$
and $G_{LT}^{\prime }(\protect\omega )$ [in MPa], according to Equations
(\protect\ref{eq4} to \protect\ref{EqH4}), computed for the mean value of
all reeds. (a): $E_{L}^{\prime }$
series A, (b): $E_{L}^{\prime }$ series B, (c): $G_{LT}^{\prime }$ series A
and (d): $G_{LT}^{\prime }$ series B. For $G_{LT}^{\prime }$, the moduli are
multiplied by 10. The abscissa of the relaxation frequencies [in Hz] is denoted by
dashed lines (series A) and dotted lines (series B). The corresponding numerical
values are listed in Tables \protect\ref{tab6} and \protect\ref{tab7}. Only
the portions of the curves between 2 and 18 $kHz$ could be fitted
adequately. The curves outside this range are purely hypothetical: we have no
measurements. }
\label{var_omega}
\end{figure}

Fig. \ref{var_omega} shows an approximation of the frequency dependence of $%
E_{L}$ and $G_{LT}$, computed for the mean value of all reeds. For Series A, the storage modulus $E^{\prime }(\omega )$ increases
from  $11700$ MPa at 2 kHz ($F1$) to $17200$ MPa at 16.8 kHz ($F4$), while for Series B, it increases from   $11100$ to $13300$ MPa. Therefore under "normal" hygrometry the reed bends  in a notably viscoelastic manner, whereas the ultra-dry reed bends in an more elastic manner.  For $G_{LT}$, the reed is generally notably viscoelastic, according to our model. The corresponding values are: 783 to 1290 MPa for Series A, and 896 to 1436 MPa for Series B. Corresponding statistics are displayed on Table \ref{tabH4}. Drying seems to increase $G_{LT}$ and decrease $E_{L}$ (except around 2300 Hz), explaining the good correlation between the variables AF1 and BF1.
\begin{table}[tbp]
\centering
\begin{tabular}{|m{30mm}|m{13mm}|m{13mm}|m{13mm}|m{13mm}|}
\hline
\textbf{Increase in}&  \multicolumn{2}{|c|}{$E_L$}&\multicolumn{2}{|c|}{$G_{LT}$} \\ \cline{2-5}
\textbf{storage modulus}& Series A & Series B& Series A & Series B \\ \hline
\textit{Mean} & $~~31\%$ & $~~16\%$ & $~~38\%$ & $~~36\%$  \\
\textit{Standard deviation} & $~~~5\%$ & $~~~6\%$ & $~~10\%$ & $~~~9\%$  \\
\textit{Minimum} & $~~20\%$ & $~~~1\%$ & $~~13\%$ & $~~15\%$  \\
\textit{Maximum} & $~~44\%$ & $~~32\%$ & $~~52\%$ & $~~49\%$  \\ \hline
\end{tabular}%
\caption{Optimal 9-coefficients model (hypothesis H4): Statistics of the frequency-dependent increase in storage modulus (for Series A or B and $E_L$ or $G_{LT}$), between the $1^{st}$ and the $4^{th}$ flexural mode: $1-E'(2 \pi f_{F1})/E'(2 \pi f_{F4})$ or $1-G'(2 \pi f_{F1})/G'(2 \pi f_{F4})$, with $f_{F1}=1996$Hz and $f_{F4}=16784$Hz. Between both series, the correlation is 0.83 for $E_L$ and 1 for $G_{LT}$ (as a consequence of the simplification of the model)}.
\label{tabH4}
\end{table}

This simplified model  permits
interesting conclusions about the structure of our data:

\begin{itemize}
\item The component $e_{1}$ is related exclusively to $G_{LT}$;

\item $e_{2}$ is related exclusively to $E_{L}$;

\item $e_{3}$ increases proportionally to the rubbery modulus  of $E_L$ (and not its glassy modulus), and decreases proportionally to the glassy modulus of $G_{LT}$ (and not its rubbery modulus); it regulates therefore the viscous component common to $E_L$ and $G_{LT}$\footnote{The parameter $E_{2}$ (respectively $G_{2}$) determines the influence of the Maxwell arm in the Zener model, since $E_{3}$ (respectively $G_{3}$) is constant in our simplified model; if $E_{2}=0$ the model is perfectly elastic and the viscous component disappears};

\item $e_{4}$ takes into account the variation in moisture content between
series A and B.
\end{itemize}

Notice that the product of the rubbery moduli for the series A, $E_{1}G_{1}$
is correlated at $72\%$ with the nominal reed strength. Individually, these
moduli correlate only at $50\%$ and $55\%$ with the nominal reed strength,
respectively.

\subsection{Results and discussion\label{results}}
Tables  \ref{tab6} and \ref{tab7} show our results for hypotheses H4 and H5. A comparison with results by other authors is difficult or even quite impossible, because of the
disparate structure of the measurements. Such a comparison requires the reconstruction of the measurement data (when possible), the fit of a viscoelastic model or the extrapolation of the values, in order to reach the frequency and temperature range of our measurements. The validity of such a highly speculative task is questionable. For the storage modulus $E_{L}^{\prime }$, all reconstructed values from other authors fall  in the range around the average of our measurements $\pm 3$ times the standard deviation, however most of the time in the lower range. This probably shows  that the selected value for the density $\rho$ is somewhat too high. No representative statistics are available by the other authors.

\begin{table}[tbp]
\centering
\begin{tabular}{|m{38mm}|m{19mm}|m{19mm}|m{19mm}|m{19mm}|}
\hline
Model &  \multicolumn{4}{|c|}{Storage Young modulus $E^{\prime}_L(\omega)$ at about $20^\circ C$} \\ \cline{2-5}
&$E_1$ [MPa] & $E_2$ [MPa] & $f_r$ [Hz] & $E^{\prime}_L$ at 4~kHz \\ \hline
Hypothesis H4, series A & 10300, SD~906 & 7309, SD~1432 & 4123, SD~808
& 13781, SD~ 897  \\ \hline
Hypothesis H5, series A & 9377, SD~844 & 8336, SD~1552 & 4027, SD~750
& 13449, SD~ 839 \\ \hline
Hypothesis H4, series B & 9423, SD~784 & 3964, SD~1109 & 2236, SD~626
& 12338, SD~ 885  \\ \hline
Hypothesis H5, series B & 7844, SD~753 & 5459, SD~1217 & 1947, SD~434
& 12168, SD~ 882 \\ \hline
\end{tabular}%
\caption{Summary of our results for hypotheses H4 and H5 about the
viscoelastic behavior of the
longitudinal Young modulus $E_L$ in cane. $E_1$, $E_2$ and $f_r$ (relaxation
frequency): parameters from Zener model. $E^{\prime}_L$ at 4~kHz : storage modulus at 4~kHz [in MPa]. SD:
standard deviation (the value preceding SD is the average). The model is valid only between 2 and 18~kHz.}
\label{tab6}
\end{table}

\begin{table}[tbp]
\centering
\begin{tabular}{|m{38mm}|m{19mm}|m{19mm}|m{19mm}|m{21mm}|}
\hline
Model & \multicolumn{4}{|c|}{Storage Shear modulus $G^{\prime}_{LT}(\omega)$ at about $20^\circ C$} \\ \cline{2-5}
& $G_1$ [MPa] & $G_2$ [MPa] & $f_r$ [Hz] & $G^{\prime}_{LT}$ at 4~kHz \\
\hline
Hypothesis H4, series A & 694,   SD~116 & 694,   SD~327 & 5420, SD~2555 &
926,   SD~102 \\ \hline
Hypothesis H5, series A & 752,   SD~119 & 628,   SD~328 & 6310, SD~3296 &
924,   SD~105 \\ \hline
Hypothesis H4, series B & 811,   SD~116 & 736,   SD~327 & 5749, SD~2555 &
1042,   SD~99 \\ \hline
Hypothesis H5, series B & 774,   SD~119 & 769,   SD~328 & 5622, SD~ 2401
& 1022,   SD~103 \\ \hline
\end{tabular}%
\caption{Summary of our results for hypotheses H4 and H5 about the
viscoelastic behavior of the shear modulus in longitudinal / tangential
plane $G_{LT}$ in cane. $G^{\prime}_{LT}$ at 4~kHz: storage modulus at 4~kHz. Same structure as Table \protect\ref{tab6}. The model is valid only between 2 and 18~kHz. }
\label{tab7}
\end{table}

The most important disagreement, compared with our model (hypotheses H4 and
H5), is about the relaxation frequency. The explanation is probably because the studied frequency range was not same. For $G_{LT}$, we found no
viscoelastic measurements by other authors. Our results are summarized in
Tables  \ref{tab6} and \ref{tab7}. For $E^{\prime}_L$, between hypotheses H4 and H5, $f_r$ and $E^{\prime}_L$ at 4~kHz agree well, whereas $E_1$ and $E_2$ diverge by about 1-2 SD. This divergence comes because the observed frequency range was not broad enough. For the shear modulus, $G_1$ and $G_2$ are in good agreement for both hypotheses (and consequently $f_r$ and $G^{\prime}_L$ at 4~kHz also).

Our model is valid only for "ambient dry" reeds (since the ultra-dry conditioning was not controlled), in a frequency range which
should not exceed one decade. We checked that a fractional derivative
model after Gaul \emph{et al.} \cite{gaul2007experimental} is not necessary in our
narrow frequency range. Such models are however really efficient to cover a
broad frequency range. For instance, the data by Lord for dry material \cite{lord2003viscoelasticity} could
be fitted very well ($E_0=8108$~MPa, $E_1=2964$~MPa, $p=0.298$~MPa$\cdot$s$^\alpha$, $\alpha=0.546$). Notice that the order of the derivative
($\alpha$) is 1 in our viscoelastic model.

Our viscoelastic model is able to partially explain  the bad correlations observed between flexural modes. In Table \ref{TableCorr}, the correlations are compared among observed resonance frequencies and computed modal frequencies, according to the viscoelastic model (hypothesis H4 and H5). It seems that additional hypotheses (such as an irregular thickness) should perhaps be considered for improving the model. However, we should remember that the determination of resonance frequencies are attached with some uncertainties, especially for the modes $F3$ and $T3$.

In order to clarify this issue, let us examine if the residuals (observed resonance frequencies \textit{minus} computed modal frequencies with hypothesis H5\footnote{Because H5 is probably less biased than H4 with its minimal structure}) contain some pertinent information. A PCA shows that perhaps 2 residual factors contain some interesting information (explaining $30\%$ and $18\%$ of the residual variance). The first residual factor is correlated with AT3 (0.86), BT2 (0.74), AT2 (0.73) and AX2 (0.66). All these modes depend strongly on $G_{LT}$. An adjustment of the coefficient $a_{m,2}$ from the metamodel for the transverse modes could probably cancel this systematic bias (remember that the coefficients $a_{m,q}$ are computed from a theoretical model, which is probably also biased). Indeed, an increase by 14, 21, 16 and  $11\%$   of this coefficient affecting the modes $T1$, $T2$, $T3$ and $T4$ makes  the $RMSD$ drop from 29.8 to 28.5 cents. The second residual factor is correlated with AF1 (0.36), AF2 (0.29), AT2 ($-0.29$), BF1 (0.28) and BT1 ($-0.23$). This probably reveals  a competition between flexural and transversal modes when fitting the model. The bias probably comes  from the coefficient $a_{m,1}$, regulating the linear dependance to $E_L$ in the metamodel. Adjusting the coefficients $a_{m,1}$ for all modes and the coefficients $a_{m,2}$ for all transverse and generic modes makes the $RMSD$ drop down to 26.2 cents, reaching  the size of the measurement steps (25 cents). The adjustments of $a_{m,1}$ are very small for the flexural modes $F1$ to $F4$: -6, 1, 6 and $1\%$.

This shows that the most important bias depends linearly on the 2 most important parameters ($E_L$ and $G_{LT}$) of our FEM computations. No supplementary parameter is required until this bias is removed (theoretically down to a $RMSD$ of 21.8 cents, according to H10). For hypothesis H4, the same linear adjustment of the metamodel lets the $RMSD$  drop from 30.4 down to 27.0 cents.

\begin{table}[tbp]
\centering

\begin{tabular}{|c|c||c|c|c||c||c|c|c|}
\hline
\textbf{Model} & \textbf{Mode} & $AF1$ & $AF2$ & $AF3$ & \textbf{Mode} & $BF1
$ & $BF2$ & $BF3$ \\ \hline
& $AF2$ & 0.73 &  &  & $BF2$ & 0.63 &  &  \\
observations & $AF3$ & 0.34 & 0.75 &  & $BF3$ & 0.43 & 0.84 &  \\
& $AF4$ & 0.28 & 0.65 & 0.74 & $BF4$ & 0.00 & 0.43 & 0.45 \\ \hline
& $AF2$ & 0.93 &  &  & $BF2$ & 0.79 &  &  \\
H4 & $AF3$ & 0.69 & 0.90 &  & $BF3$ & 0.62 & 0.97 &  \\
& $AF4$ & 0.56 & 0.81 & 0.98 & $BF4$ & 0.36 & 0.82 & 0.91 \\ \hline
& $AF2$ & 0.90 &  &  & $BF2$ & 0.69 &  &  \\
H5 & $AF3$ & 0.57 & 0.87 &  & $BF3$ & 0.52 & 0.98 &  \\
& $AF4$ & 0.43 & 0.77 & 0.98 & $BF4$ & 0.26 & 0.84 & 0.92 \\ \hline
\end{tabular}
\caption{Correlations of resonance frequencies between flexural modes within the same series, after observations and viscoelastic models H4 and H5. $AF1$, $AF2$, $AF3$, $BF1$, $BF2$ and $BF3$: correlations with modal frequencies $F1$, $F2$ and $F3$ within series A or series B.  Lines:  corresponding flexural mode between which the correlations are computed.
}
\label{TableCorr}
\end{table}

\section{Conclusion}\label{Conclusion}

The numerical model is satisfactory. From the statistical analysis are discussed in Section \ref{ConclStat}, it allows  selecting the most important parameters describing the mechanical behavior of a reed.

 The efficient elastic metamodel can be extended to a viscoelastic behavior of the reeds,
approximating the resonance frequencies from the longitudinal Young modulus $E_{L}$ and the longitudinal / transverse shear modulus $G_{LT}$ and  considering the hypothesis of their frequency dependence. A reconstruction of the observed resonance frequencies can  be achieved with a
good accuracy, estimating for each reed only 4 components, from which the parameters of a viscoelastic model are computed as a linear combination. The selected model (according to hypothesis H4) is probably slightly biased, but it is more robust against measurement errors than more refined models.

Table \ref{Table_table_2} shows that these components are highly correlated
to the factors computed by PCA (0.96 to 0.98).

The proposed method allows the determination of 3 mechanical parameters characterizing the material composing each reed, with a single series of measurements, using Equations (\ref{eq1}, \ref{eq4} and \ref{EqH4}). The reed should be conditioned with a relative humidity corresponding to the one ensured by the hermetically sealed package by Vandoren (about $55\%$). The fourth parameter cannot be determined in a reproducible way, since the exposure of the reeds to the ultra-dry air of the optical laboratory was not controlled. The same protocol and the same viscoelastic model can be used for other kinds of single reeds (bass clarinet, saxophone). Only the coefficients of Table \ref{tab9} have to be recomputed after a FEM simulation of the corresponding reed shape.

Despite the fact that the eigenmodes of higher order probably play no important role
in the acoustics of the clarinet, the present study shows that they reveal the inner
structure of the material building the tip of the reed, so a new step could be done for an objective mechanical characterization of the clarinet reed.
A subsequent study should examine if the obtained components are correlated with some musical qualities of the reeds. This could help the reed makers to gain a better control on their products.

\bigskip \textbf{Acknowledgments} \bigskip

We express gratitude for the financial aid from association DEPHY of the
"département de physique de l'ENS", and to Fadwa Joud for her help during the
experiments. The Haute-Ecole ARC (Neuchâtel, Berne, Jura) provided the software for
simulation and calculation, and offered helpful instructions. The
Conservatoire neuchâtelois partially supported this study for nearly 3
years. We also thank Bruno Gazengel, Morvan Ouisse and Emmanuel Foltête for fruitful
discussions, and the referees for useful suggestions.

\begin{center}
\LARGE{APPENDICES}
\end{center}

\appendix
{\label{Appendix}}
\section{Statistics and correlations}
\label{statistic}
The table \ref{Table_table_1} gives the detailed statistics.
\begin{table}[tbp]
\centering
\begin{tabular}{|c||c|c|c|c|c||c|c|c|c|c|c|}
\hline
& \multicolumn{5}{c||}{Series A} & \multicolumn{5}{c|}{Series B}&   \\ \hline
Mode & \textbf{\ }$N_{A}$ & $\mu $ & $\sigma $ & $\mathbf{\min }$ & $\mathbf{%
\max }$ & \textbf{\ }$N_{B}$ & $\mu $ & $\sigma $ & $\mathbf{\min }$ & $\mathbf{%
\max }$  & \\ \hline
F1 & 55 & 1996 & 77 & 1838 & 2154 & 55 & 1960 & 66 & 1812 & 2093 &  \\
T1 & 55 & 3377 & 132 & 3091 & 3676 & 55 & 3436 & 129 & 3136 & 3729 &  \\
F2 & 55 & 5130 & 161 & 4767 & 5669 & 55 & 4856 & 189 & 4435 & 5351 &  \\
T2 & 55 & 6108 & 261 & 5587 & 6939 & 55 & 6193 & 247 & 5669 & 7040 &  \\
X1 & 55 & 6869 & 198 & 6455 & 7458 & 45 & 6801 & 217 & 6363 & 7458 &  \\
T3 & 55 & 9571 & 500 & 8617 & 11014 & 54 & 9590 & 458 & 8742 & 11014&   \\
F3 & 55 & 10146 & 419 & 9262 & 10857 & 55 & 9213 & 414 & 8372 & 10396 &  \\
X2 & 55 & 11521 & 387 & 10701 & 12187 & 54 & 11688 & 379 & 10857 & 13098 & \\
X3 & 55 & 12294 & 368 & 11502 & 13482 & 11 & 12290 & 471 & 11839 & 13482 &
\\
T4 & 53 & 14011 & 756 & 12186 & 16503 & 54 & 14111 & 763 & 12186 & 16503 &
\\
F4 & 45 & 16784 & 552 & 15803 & 18524 & 54 & 15363 & 663 & 14079 & 17484 &
\\
X6 & 41 & 16984 & 972 & 15133 & 18793 & 30 & 16888 & 734 & 15577 & 18258 &
\\
X4 & 24 & 18497 & 518 & 17234 & 19067 & 23 & 17544 & 772 & 16033 & 19911 &
\\
X5 & 0 &  &  &  &  & 54 & 18896 & 501 & 17484 & 19911  & \\
T5 & 0 &  &  &  &  & 14 & 19668 & 329 & 19067 & 19911  & \\ \hline
\textbf{Total} & 658 &  &  &  &  & 668 &  &  &  &  &  \\ \hline
\end{tabular}%
\caption{Observed resonance frequencies, sorted by frequency (in Hz). $N_{A}$
(resp. $N_{B}$): Number of identified pattern for each mode of series A
(resp. B). $\protect\mu $: Mean value of the resonance frequency, $\protect%
\sigma $: Standard deviation, $\mathbf{\min }$: Minimum, $\mathbf{\max }$:
Maximum. \textbf{Total}: total number of identified patterns for each
series. }
\label{Table_table_1}
\end{table}

Linear correlations have been computed between all possible couples of
variables in a usual way (all the variables from Table \ref{Table_table_1},
except $X5$ and $T5$ of series A, i.e. 13 variables for series A and 15 for
series B, plus the nominal reed strength). Results of series A
for the mode $F1$ are denoted AF1, and similarly for the other results. The
following 12 pairs have a correlation greater than 0.9: AF1/BF1\footnote{%
AF1/BF1 means AF1 \emph{versus} BF1. The correlations are computed between $%
AF1[n]$ and $BF1[n]$, for $n=1$ to $N$, where $N=55$; missing observations are deleted.}, AT1/BT1, AT2/BT2, AT3/AT4, AT3/BT3,
AT3/BT4, BT3/BT4, AT4/BT3, BX1/BX3, AT4/BT4, BT4/BX6 and BX3/BX4. 54 pairs
of variables have a correlation between 0.8 and 0.9, 50 other pairs between
0.7 and 0.8 and 262 other pairs, below 0.7.

\emph{Between the two series}, the correlation is excellent for corresponding
transversal modes ($T1$ (i.e. AT1/BT1): 0.97, $T2$: 0.97, $T3$: 0.96 and $T4$%
: 0.98), and generally good for corresponding generic modes ($X1$: 0.87, $X2$%
: 0.84, $X3$: 0.87 and $X4$: 0.55). For flexural modes, the correlation is
good for $F1$ and progressively lower for increasing mode order ($F1$: 0.92, $F2$: 0.66,
$F3$: 0.57 and $F4$: 0.47).

\emph{Within the same series}, on the contrary, there is a poor correlation between AF1 and all measurements
of series A, and similarly for BF1 and series B. This is striking: the two
best correlated variables are AF2 and AX1 (0.73 and 0.49, respectively) for
AF1 and BF2 and BT1 (0.63 and 0.49, respectively) for BF1. Moreover, these correlations are quite low  among all flexural modes: see Table \ref{TableCorr}. This fact is discussed in section \ref{results}.

The nominal reed strength correlates at 0.7 with AT1 and AX4. We expected a
better correlation with $F1$ (only 0.6). This is surprising, since the reeds
were probably sorted by a quasi-static bending method by the manufacturer.
This would mean that the storage modulus of $E_{L}$ at very low frequency is
not well correlated with its value at the frequencies of the measured
resonances. The influence of density has also to be considered. However, Obataya \emph{et al.} \cite{obataya1999a_effects} observed a good correlation between density and $E_L$. This point has to be investigated (see also section \ref%
{simplifiedModel} in Appendix \ref{DVM}).

\section{Defining the shape of the reed}

\label{shape}
\newcolumntype{M}[1]{>{\raggedleft}m{#1}}
\begin{table}[h]
  \centering
  \begin{tabular}{|r|M{14mm}|M{14mm}|M{14mm}|}
    \hline
x [mm]&$s_0$ [mm] y=0 mm&$s_1$ [mm] y=4mm&$s_2$ [mm]
y=6mm\tabularnewline \hline 0&     0.074 &     0.080 &
0.042\tabularnewline 5&     0.343 &     0.293 &
0.197\tabularnewline 10&     0.648 &     0.542 &
0.377\tabularnewline 15&     1.047 &     0.847 &
0.571\tabularnewline 20&     1.451 &     1.135 &
0.745\tabularnewline 25&     1.926 &     1.527 &
1.078\tabularnewline 30&     2.540 &     2.084 &
1.589\tabularnewline 35&     3.351 &     2.817 &
2.256\tabularnewline

    \hline
  \end{tabular}
  \caption{Network of points for interpolating the thickness of the vamp. See explanations in the text.}\label{tab8}
\end{table}

The thickness of the vamp at point $(x,y)$ is interpolated as follows:
first, we interpolate 3 points at $y$=0, 4 and 6mm, with 3 cubic
splines, according to Table \ref{tab8}. These three points ($s_{0}(x)$, $s_{1}(x)$ and $s_{2}(x)$) define a biquadratic
polynomial:
\begin{equation*}
vamp(x,y)=p_{0}(x)+p_{1}(x) \, y^2+p_{2}(x) \, y^4
\end{equation*}
with $p_{0}(x) = s_{0}(x)$, $p_{1}(x) = (-65\, s_{0}(x) + 81\, s_{1}(x) - 16\, s_{2}(x))/720$ and $p_{2}(x) =
(5\, s_{0}(x) - 9\, s_{1}(x) + 4\, s_{2}(x))/2880$, allowing an interpolation on the $y$
axis. The network of points above was estimated using a least
squares fit, based on a network of $12 \times 24$ thickness measurements,
achieved with a dial indicator and a coordinates-measuring table (estimated
accuracy: $\pm 5 \mu m$ in $z$, $\pm 50 \mu m$ in $x$ and $y$). We measured
twenty reeds and select a particularly symmetrical one as reference. This
method allows the reconstruction of the measurement network with an accuracy
of $\pm 10 \mu m$.

The thickness of the heel is defined to be :
\begin{equation*}
heel(y)=-14.1+\sqrt{17.4^2-y^2}
\end{equation*}

The contour of the reed in the xy-plane is defined by:
\begin{equation*}
contour(x)=
\begin{cases}
 0 & x<0 \text{ or }x\geq 67.5\\
 \sqrt{(24.4-x) x} & x<1.13196 \\
 4.08044+\sqrt{-5.31+6.8 x-x^2} & x<2.94661 \\
 \frac{263}{40}-\frac{11}{900} x & x<67.5
\end{cases}
\end{equation*}

The thickness of the reed at point $(x,y)$ is defined by:
\begin{equation*}
thickness(x,y)=
\begin{cases}
 min[heel(y), vamp(x,y)] & Abs(y)< contour(x) \\
 0 & \text{otherwise}
\end{cases}
\end{equation*}

\section{Coefficients of the metamodel}

\label{meta}
\begin{table}[h!]
\centering
\begin{tabular}{|c|c|r|r|r|r|r|r|r|r|}
\hline
mode & m & $a_{m,0}$ & $a_{m,1}$ & $a_{m,2}$ & $a_{m,3}$ & $a_{m,4}$ & $a_{m,5}$ &
$\delta -$ & $\delta +$ \\ \hline
$F1$ & 1 & 2334.56 & -1165877 & 0.2763 & -21.99 & 145883795 & -0.000324 & -2 &
4 \\
$T1$ & 2 & 1481.70 & -642027 & 5.0725 & 569.69 & 83451520 & -0.005886 & -3 & 3
\\
$F2$ & 3 & 3651.79 & -1060625 & 0.7082 & -123.77 & 130061359 & -0.000786 & -2
& 5 \\
$T2$ & 4 & 2403.82 & -493591 & 3.8130 & 501.70 & 64716271 & -0.003741 & -9 & 5
\\
$X1$ & 5 & 3153.73 & -822597 & 3.1425 & 604.29 & 105032670 & -0.003785 & -3 & 4
\\
$F3$ & 6 & 4669.87 & -1009348 & 0.7623 & -101.18 & 122741172 & -0.000877 & -3
& 6 \\
$T3$ & 7 & 3015.77 & -275531 & 3.5285 & 250.00 & 29933670 & -0.003016 & -2 & 1
\\
$X2$ & 8 & 3874.44 & -633268 & 2.0921 & 589.25 & 83462866 & -0.001958 & -9 & 6
\\
$X3$ & 9 & 4381.85 & -926543 & 1.9907 & 730.31 & 122925338 & -0.002593 & -3 & 3
\\
$T4$ & 10 & 2659.08 & 588457 & 6.6088 & -1269.38 & -120493740 & -0.005269 & -19
& 26 \\
$F4$ & 11 & 6450.01 & -1689363 & -2.5544 & 1282.75 & 247738162 & 0.001631 & -32
& 42 \\ \hline
\end{tabular}%
\caption{Coefficients of the metamodel, Equation (\protect\ref{eq1}). The maximum negative and
positive deviations of the model (compared to the values calculated by FEM)
are given by $\protect\delta -$ and $\protect\delta +$ [in cents]. The mode $L1$ (6th
mode) has been deleted from the Table, since we didn't observe it.}
\label{tab9}
\end{table}
\clearpage
\section{Development and selection of a simplified viscoelastic model}
\label{DVM}
In this section we describe how the proposed model was developed and selected. Alternate options are presented.
\subsection{Data structure}\label{dataStruc}

We need a specific notation for denoting our complicated multivariate data
structure as arrays, after a list of indices (see Table \ref{tabIndices}).
The 11 modes are defined after Table \ref{tab9} (from 1
to 11: $F1$, $T1$, $F2$, $T2$, $X1$, $T3$, $F3$, $X2$, $X3$, $T4$, $F4$).
\begin{table}[tbp]
\centering%
\begin{tabular}{|c|c|c|l|}
\hline
Indice & from & to & numbering \\ \hline
n & 1 & N=55 & reeds \\
s & 1 & S=2 & series of measurements A (s=1) and B (s=2) \\
m & 1 & M=11 & modes \\
q & 0 & Q=5 & coefficients for Equation (\ref{eq1}) \\
i & 1 & I=2 & moduli $E_{L}$ (i=1) or $G_{LT}$ (i=2) \\
j & 1 & J=3 & viscoelastic parameters \\
k & 0 & K=2,3 or 4 or 12& components (factors) \\ \hline
\end{tabular}%
\caption{List of indices}
\label{tabIndices}
\end{table}
We define a variable $v_{n,s,i,j}$ holding all parameters of our
viscoelastic model:
\begin{equation}
v_{n,s,1,j}=E_{j}\text{ and }v_{n,s,2,j}=G_{j}\text{ for reed }n\text{ and
series }s.  \label{vvv}
\end{equation}

The array $r_{n,s,m}$ holds the reconstructed resonance frequencies computed for all reed, series and modes,
with the parameters $v_{n,s,i,j}$ of our viscoelastic model (see Section \ref{ViMo}).

\subsection{Mean Squared Deviation}\label{MSD}
As a cost function to minimize, we define the
Mean Squared Deviation MSD (also called Mean Squared Error) between reconstructed and measured resonance
frequencies $o_{n,s,m}$:
\begin{equation}
MSD=\frac{1}{NSM}\sum_{n=1}^{N}\sum_{s=1}^{S}%
\sum_{m=1}^{M}(o_{nsm}-r_{nsm})^{2}  \label{eq5d}
\end{equation}%
With Eq. (\ref{eq5d}), the components of array $\mathbf{v}$ can be
fitted by any appropriate algorithm for minimizing a multivariate function.
All available measurements can be utilized for fitting the models\footnote{%
This was not the case with PCA.}. Missing observations $o_{m,s,n}$ are then
eliminated while computing $MSD$.

Our model allows a quite good reconstruction of the resonance frequencies, with
a $\sqrt{MSD}\equiv RMSD$ (Root Mean Square Deviation) smaller than 20 cents (it is very small for lower modes, and always
smaller than 25 cents). This corresponds to the hypothesis named H9 (all hypotheses H1 to H11 are presented and commented in Section \ref{Hypot}), but, as explained in section \ref{invProb},  the values of the coefficients are not always
plausible physically.  In Section \ref{viscomodel}, we examine how to regulate this
drawback by multiple regression.

\subsection{Estimating the parameters of the
viscoelastic model by multiple regression\label{viscomodel}}

Multivariate linear regression consists in projecting linearly a $n$
dimensional space on a onedimensional space. The generic equation is
\begin{equation}
y=a_{0}+\sum_{n=1}^{N}a_{n}x_{n}=\sum_{n=0}^{N}a_{n}x_{n},\quad \text{if } x_{0}=1  \label{eq5e}
\end{equation}%
Equation (\ref{eq5e}) can be generalized for multiple regression as:
\begin{equation}
\mathbf{y}=\mathbf{Ax}  \label{eq5g}
\end{equation}%
where $\mathbf{x}$\textbf{\ }and $\mathbf{y}$ are column vectors and $%
\mathbf{A}$ a matrix. In what follows, in order to use conventional  vectors and matrices
(with respectively one and two dimensions), we use the following notation: $%
v_{j}[n,s,i]=v_{n,s,i,j}$ \ is the $j$th element of the column vector $%
\mathbf{v}[n,s,i].$

In our case, we have no prior knowledge about the relationships among the
parameters of our model. We
assume that each parameter in the model can be computed as a linear
combination of some unknown independent components by multiple regression. The
multiple regression formula can be written as:%
\begin{equation}
v_{j}[n,s,i]=\overset{K}{\underset{k=0}{\sum }}M_{jk}[s,i]\cdot e_{k}[n]
\label{eq.LRE}
\end{equation}

In conventional vectors and matrices notation, Eq. (\ref{eq.LRE}) reads:
\begin{equation*}
\boldsymbol{v}[n,s,i]=\boldsymbol{M}[s,i]\cdot \boldsymbol{e}[n]
\end{equation*}

where $\boldsymbol{e}[n]$ is the vector of the orthogonal components for each observed reed $n$%
, $\boldsymbol{M}[s,i]$ is the regression matrix, independent of the reed
number, depending on the series and the kind of modulus ($E_{L}$ or $G_{LT}$%
). $\boldsymbol{v}[n,s,i]$ is the vector of the parameters of the viscoelastic model.

We have only to choose an arbitrary number of components, for instance $K=4$,
referring to our PCA. We introduce arbitrary constraints in order to obtain
more comparable results: over the different reeds, we state that the
components must be orthogonally normalized (mean 0, standard deviation 1 and intercorrelation 0). The matrix of components $\boldsymbol{e}$ has consequently to satisfy:

\begin{equation}
\boldsymbol{e}\cdot \boldsymbol{e}^{T}=\left(
\begin{array}{ccccc}
N & 0 & 0 & 0 & 0 \\
0 & N-1 & 0 & 0 & 0 \\
0 & 0 & N-1 & 0 & 0 \\
0 & 0 & 0 & N-1 & 0 \\
0 & 0 & 0 & 0 & N-1 %
\end{array}%
\right)
\label{eq.ortho}
\end{equation}

Each individual column vector $\mathbf{e}[n]$ from this matrix is written as
follows (for the component $e_{0}[n]=1$ : see Equation (\ref{eq5e})):%
\begin{equation}
\boldsymbol{e}[n]=\left(
\begin{array}{ccccc}
1 & e_{1}[n] & e_{2}[n] & e_{3}[n] & e_{4}[n]%
\end{array}%
\right) ^{T}.
\label{eqVe}
\end{equation}

The components of $\boldsymbol{M}$\textbf{\ }are fitted by minimizing $MSD$ in
Eq. (\ref{eq5d}). As starting value for the orthogonal components we set : $%
e_k[n]=factor_k[n]$, where $factor_k[n]$ are the factors computed by PCA (see Section\ref{pcaDefinition}).
After a first
estimation of $\boldsymbol{M}$, it is possible to release the approximation
about $\boldsymbol{e}$: all components and all coefficients in the matrices can be fitted by
the fitting procedure. However the number of variables to fit is probably
much higher than allowed by the most algorithms of function minimizing. The
fitting procedure has to be carried "by hand" with subsets of variables. The
procedure we used is described below.

For $K=4$ (the dimension of vector $\boldsymbol{e}[n]$ being
5), and some supplementary choices, the model works very well.

\subsection{Empirical simplification of a 4-parameter model}

\label{simplifiedModel}

We observed that the effect of hygrometric changes between both measurements series can be taken into account with one parameter only, practically without drop in quality of fit. This effect can be isolated on component $e_4[n]$ and the remaining components can be transformed linearly, so the further computations can be achieved  from a common basis. Eq. (\ref{eq.LRE}) is then structured as:
\begin{equation}
\boldsymbol{M}[s,i]=\boldsymbol{\hat{M}}[i]\cdot \boldsymbol{\check{M}}[s]\text{ ,}
\label{eq5k}
\end{equation}%
The reasoning considers the situation where the two series of observations
are independent, with a number of components in vector $\boldsymbol{\check{e}}%
[s,n]=\boldsymbol{\check{M}}[s]\cdot \boldsymbol{e}[n]$ reduced to $\check{K}%
=K-1=3 $. As a consequence the number of rows of matrix $\boldsymbol{\check{M%
}}[s]$ is 4, as well as the number of columns of $\boldsymbol{\hat{M}}[i]$.\
This approach, which allows separating hygrometry effects, offers a comfortable
way to test different hypotheses, without changing the structure of the
computation, by setting some coefficients in the matrices at some arbitrary
values or by introducing some linear relationship between coefficients. We
get fewer "active" coefficients to fit in the model: the fitting procedure
is much faster and this raises the probability to find the best possible fit.

We tried to minimize the number of coefficients different from zero in the
matrices, without substantial drop in quality of fit. The model was fitted
"by hand", using the solver of Excel (Microsoft Office). We found
empirically that a quite sparse setup still gives a good fit\footnote{%
The fitting process was realized by repeatedly performing four kinds of procedures, in
an arbitrary order :
\par
\begin{enumerate}
\item Fit the active coefficients in model, without adjusting the
coefficients $a_{m,0}$%
\par
\item Fit the active coefficients in model and adjust the coefficients $%
a_{m,0}$ (or fit all coefficients: $a_{m,q}$).
\par
\item Fit the $e_{n,k}$ components (individually for each reed, irrespective of Equation (\ref{eq.ortho})), then
then normalize and orthogonalize components (in order to satisfy Eq. (\ref{eq.ortho})), finally rotate components (for
improving the fit).
\par
\item Eliminate some active coefficients (set them as 0, 1 or some other
constant value; set some arbitrary linear dependence from other coefficients)
\end{enumerate}
}:
\begin{equation}
\boldsymbol{\hat{M}}[EL]=\left(
\begin{array}{cccc}
d_{10} & 0 & d_{12} & d_{13} \\
d_{20} & 0 & d_{22} & d_{23} \\
d_{30} & 0 & 0 & 0%
\end{array}%
\right) \text{ \ \ ; \ }\boldsymbol{\hat{M}}[GLT]=\left(
\begin{array}{cccc}
d_{40} & d_{41} & 0 & 0 \\
d_{50} & d_{51} & 0 & d_{53} \\
d_{60} & 0 & 0 & 0%
\end{array}%
\right)  \label{eq9}
\end{equation}%
\begin{equation}
\boldsymbol{\check{M}}[SeriesA]=\left(
\begin{array}{ccccc}
1 & 0 & 0 & 0 & 0 \\
0 & 1 & 0 & 0 & 0 \\
0 & 0 & 1 & 0 & 0 \\
0 & 0 & 0 & 1 & 0%
\end{array}%
\right) \text{ \ \ ; \ }\boldsymbol{\check{M}}[SeriesB]=\left(
\begin{array}{ccccc}
1 & 0 & 0 & 0 & 0 \\
c_{10} & 1 & 0 & 0 & 0 \\
c_{20} & 0 & c_{22} & 0 & c_{24} \\
c_{30} & 0 & 0 & 1 & 0%
\end{array}%
\right)  \label{eq10}
\end{equation}%
This corresponds to the hypothesis named H5.
Furthermore some coefficients may be proportional to others, without
noticeable drop in quality of fit. This diminishes the number of active
coefficients in the different matrices from 18 to 9 (hypothesis H4):

\begin{eqnarray}
d_{22} &=&2d_{12}\text{ ; }d_{13}=d_{12}=-\text{ }d_{23}\text{;}  \notag \\
d_{40} &=&d_{50}=6d_{41}\text{ ; }d_{51}=2d_{41}=-d_{53}\text{ ;}  \label{11}
\\
c_{22} &=&c_{24}=1/2.  \notag
\end{eqnarray}%

\subsection{Adjusting coefficients for removing systematic errors}
After fitting the different models, we observed some systematic deviations
in the resonance frequencies between model and observations. This error has
probably two different origins: an inevitable inaccuracy in the FEM
computation (and in our metamodel) and an error for parameters
not included in the model. A straightforward way to minimize
the residuals is to fit the coefficients $a_{m,0}$ in Eq. (\ref%
{eq1})\footnote{%
These coefficients can be fitted through the fitting procedure (some constraints have however to be introduced, to avoid an important deviation  from their theoretical values) or merely
adjusted \textit{a posteriori}, so that the total averaged deviation for each mode and both series
is 0.}. Fitting all coefficients in
Eq. (\ref{eq1}) is doubtless a more questionable way to reduce this
error (hypotheses H8 and H9). This can reduce
the mean deviation between model and observations, but greatly increase the
number of coefficients in the model (see however the discussion in Section\ref{results}).

\subsection{Testing different hypotheses \label{Hypot}}

We tested different hypotheses with our model, in order to select a
particularly efficient model. Some of them are summarized in the Table \ref%
{tab4a}.
\begin{table}[tbp]
\centering
\begin{tabular}{|c|c|c|c|c|c|}
\hline
\textbf{Hypothesis} & \textbf{$K$} & \textbf{model} & \textbf{$\#$ActiveCoef} & \textbf{$\#$OtherCoef}& \textbf{\textit{RMSD}} \\ \hline
H1 & 0 & elastic & 2 & $a_{m,0}\rightarrow 11$ & 76.2\\ \hline
H2 & 2 & elastic & 4 & $a_{m,0}\rightarrow 11$ & 54.8 \\ \hline
H3 & 3 & elastic & 7 & $a_{m,0}\rightarrow 11$ & 43.9 \\ \hline
H4 & 4 & viscoelastic & 9 & $a_{m,0}\rightarrow 11$ & 30.4 \\ \hline
H5 & 4 & viscoelastic & 18 & $a_{m,0}\rightarrow 11$ & 29.8 \\ \hline
H6 & 4 & viscoelastic & 44 & $a_{m,0}\rightarrow 11$ & 29.1 \\ \hline
H7 & 4 & viscoelastic & 60 & $a_{m,0}\rightarrow 11$ & 28.6 \\ \hline
H8 & 4 & viscoelastic & 60 & $a_{m,q}\rightarrow 66$ & 23.2 \\ \hline
H9 & 12 & viscoelastic & - & $a_{m,q}\rightarrow 66$ & 19.8 \\ \hline
H10 & 4 & regression & - & $W_{s,m,k}\rightarrow 110$ & 21.8 \\ \hline
H11 & 4 & regression & 4 & $\hat W_{m,\check{k}}\rightarrow 44$ & 24.7 \\ \hline
\end{tabular}%
\caption{Synthesis of some hypotheses tested with our model. \textbf{$K$}: Number of components for each reed $n$; \textbf{model}: elastic, viscoelastic or multiple regression (with the elastic model, $E_L$ and $G_{LT}$ are independent from frequency; the elastic model is computed after the viscoelastic one by setting very small values for the coefficients affecting $E_2$, $%
E_3$, $G_2$ and $G_3$, in order to
avoid division by 0; multiple regression: see Equations (\protect\ref{eq14a} and \protect
\ref{eq15a})); \textbf{$\#$ActiveCoef}: number of active coefficients in the matrices $\boldsymbol{M}$ or $\boldsymbol{\hat{M}}$ and $\boldsymbol{\check{M}}$ (fitted through the fitting procedure); \textbf{$\#$OtherCoef}: number of other coefficients estimated in model ($a_{m,0}$: adjusted so that the mean error for each mode $m$ and both series is zero, $W_{s,m,k}$: computed analytically, otherwise: fitted through the fitting procedure); \textbf{\textit{RMSD}} [in cents] after Equation (\protect\ref{eq5d}):
this is a measure of goodness of fit (remember that the vibration patterns
of the reeds were observed in steps by 25 cents).}
\label{tab4a}
\end{table}

Hypotheses H1 to H3 (elastic model) present a poor fit; the adjustments for coefficients
$a_{m,0}$ are large, compensating partially for the missing viscoelastic
components. All viscoelastic models are notably better and exhibit smaller
adjustments for $a_{m,0}$. Frequency dependence for $E_L$ and $G_{LT}$ seems
evident. Hypothesis H4 shows a good accuracy, with only 9 fitted
coefficients (and 11 adjustments). Increasing the number of coefficients up to
60 brings only a marginal contribution (H5 to H7). Adjusting the other
coefficients of Equation~(\ref{eq1}) in H8 and H9 improve the model, especially for
the higher modes (Notice that no multiple regression is used for H9). Our FEM computations (and consequently our metamodel) are
probably attached with systematic errors in this frequency range. The
influence of $E_T$ should possibly be considered. Between H8 and H9, the
total number of components ($N\times K$) increases from 220 to 660. The adjustments within morphological classes are related: \textquotedblleft flexural\textquotedblright\ modes shows systematically lower values than neighboring \textquotedblleft transversal\textquotedblright\ modes.

\subsection{Backward random validation}
Following a suggestion by a referee, we applied a principal component analysis to simulated data computed after
hypothesis H5: we assigned randomly a value for the 4 components and 55
reeds, following a normal distribution. We repeated this operation ten
times. As expected, the PCA detected 4 factors capturing $91.2\%$
(Standard Deviation $1.3\%$) of the variance of the simulated data (mean:
42.6, 20.9, 16.2 and $11.5\%$ for each factor). This seems compatible with the
analysis performed on the observed frequencies (Section \ref{pcaDefinition}): 4
factors: $91.2\%$ of the variance (53.6, 21.4, 10.8 and $5.4\%$ for each
factor).

\section{Reconstructing observed resonance frequencies by multiple regression}\label{reconstr}

The scheme of our viscoelastic model is: \\$\langle$vector $%
\boldsymbol{e}\rangle\rightarrow\{$multiple regression \textsf{Equation (\ref{eq.LRE})}$\}\rightarrow \langle$viscoelastic
coefficients $\boldsymbol{v}\rangle\rightarrow\{$viscoelastic model \textsf{Equation (\ref{eq4})}$\}\rightarrow \langle$%
moduli $E_L^{\prime}$ and $G_{LT}^{\prime}\rangle\rightarrow\{$metamodel
\textsf{Equation (\ref{eq1})}$\}\rightarrow \langle $reconstructed resonance frequencies $\boldsymbol{r}\rangle$.

It has a very interesting property: the same model of cane can be used for any
kind of reeds (for instance bass clarinet or saxophone) or for any other
boundary conditions. Only Equation (\ref{eq1}) has to be changed (or at least, the coefficients $a_{m,q}$ have to be recomputed).

Within our particular setup, the viscoelastic model is however not required
for reconstructing the observed resonance frequencies: PCA is theoretically the
optimal linear scheme, in terms of least mean square error, for compressing
a set of high dimensional vectors into a set of lower dimensional vectors
and then reconstructing the original set by multiple regression. The
shortened scheme is merely: \\$\langle$vector $%
\boldsymbol{factor}\rangle\rightarrow\{$multiple regression$\}\rightarrow \langle$reconstructed resonance
frequencies $\boldsymbol{r}\rangle$.

Let us examine this option. For this purpose we use the array $factor_{n,k}$ computed in Section \ref{pcaDefinition}%
, holding our 4 principal components (factors). As before (Section \ref{viscomodel}), we set $factor_{n,0}=1$%
. The array of reconstructed resonance frequencies $r_{n,s,m}$ can be computed
by multiple regression using an array of matrices $W_{s,m,k}$:
\begin{equation}  \label{eq14a}
\boldsymbol{r}[n,s]=\boldsymbol{W}[s] \cdot \boldsymbol{factor}[n]
\end{equation}
As before (Section \ref{simplifiedModel}), we have also the option to reduce the dimensionality from $K$ to $%
\check K$, using the previously defined array of matrices $\boldsymbol{\check M}[s]$ and
then use a unique matrix $\hat W_{m,\check k}$ to operate the multiple
regression:
\begin{equation}  \label{eq15a}
\boldsymbol{r}[n,s]=\boldsymbol{\hat W} \cdot \boldsymbol{\check M}[s]\cdot \boldsymbol{factor}[n]
\end{equation}

We call these 2 options: hypotheses H10 and H11. For H11, we performed a
small orthogonal rotation of the factors to concentrate the information
about the hygrometric material properties in $factor_{n,4}$, for a better fit. The
results are summarized in Table \ref{tab4a}.

Multiple regression is an accurate way to retrieve our measurements, comparable
to H8 and H9. With only 48 coefficients, H11 is very efficient, even better for
\textquotedblleft transversal modes\textquotedblright\ as H10. The results
of the regressive model are more difficult to interpret than those of the
viscoelastic model. As $e_{n,4}$ before, $factor_{n,4}$ serves uniquely to adjust $%
factor_{n,2}$ relatively to series B. As expected, $factor_{n,1}$ influences mainly the
\textquotedblleft transversal modes\textquotedblright\ and $factor_{n,2}$ the
\textquotedblleft flexural modes\textquotedblright . Their respective
coefficients in the matrix $\boldsymbol{\hat{W}}$ reflect this antinomy.
\textquotedblleft Transversal\textquotedblright\ and \textquotedblleft
flexural modes\textquotedblright\ are concerned by $factor_{n,3}$ in a quite
similar way, but the slope is not same.

\bibliographystyle{plain}
\bibliography{taillard}

\begin{thebibliography}{10}

\bibitem{spatz1997biomechanics}
H.C. Spatz, H.~Beismann, F.~Brüchert, A.~Emanns, and T.~Speck.
\newblock {Biomechanics of the giant reed Arundo donax}.
\newblock {\em Philosophical Transactions of the Royal Society of London.
  Series B: Biological Sciences}, 352(1349):1, 1997.

\bibitem{marandas1994caractérisation}
E.~Marandas, V.~Gibiat, C.~Besnainou, and N.~Grand.
\newblock {Caract{\'e}risation m{\'e}canique des anches simples d'instruments
  {\`a} vent}.
\newblock {\em J. Phys. IV France}, 4:C5--633, 1994.

\bibitem{Ollivier_2002}
S.~Ollivier.
\newblock {\em Contribution à l'étude des oscillations des instruments à vent à
  anche simple: Validation d'un modèle élémentaire}.
\newblock PhD thesis, Université du Maine, Le Mans, France., 2002.

\bibitem{dalmont2003nonlinear}
J.P. Dalmont, J.~Gilbert, and S.~Ollivier.
\newblock {Nonlinear characteristics of single-reed instruments: Quasistatic
  volume flow and reed opening measurements}.
\newblock {\em The Journal of the Acoustical Society of America}, 114:2253,
  2003.

\bibitem{speck2003mechanical}
O.~Speck and HC~Spatz.
\newblock {Mechanical Properties of the Rhizome of Arundo donax L.}
\newblock {\em Plant biol (Stuttg)}, 5:661--669, 2003.

\bibitem{speck2004damped}
O.~Speck and H.C. Spatz.
\newblock {Damped oscillations of the giant reed Arundo donax (Poaceae)}.
\newblock {\em American journal of botany}, 91(6):789, 2004.

\bibitem{Chevaux_1997}
Ph. Chevaux.
\newblock Améliorations de la durée de vie des anches doubles en roseau pour
  instruments à vents.
\newblock Projet de fin d'étude, Institut National des Sciences appliquées de
  Lyon, France, 1997.

\bibitem{obataya1996_physical}
E.~Obataya.
\newblock {Physical properties of cane used for clarinet reed}.
\newblock {\em Wood Res. Tech. Notes}, 32:30--65, 1996.

\bibitem{obataya1999a_effects}
E.~Obataya, T.~Umezawa, F.~Nakatsubo, and M.~Norimoto.
\newblock {The effects of water soluble extractives on the acoustic properties
  of reed (Arundo donax L.)}.
\newblock {\em Holzforschung}, 53(1):63--67, 1999.

\bibitem{obataya1999b_mechanical}
E.~Obataya and M.~Norimoto.
\newblock {Mechanical relaxation processes due to sugars in cane (Arundo donax
  L.)}.
\newblock {\em Journal of Wood Science}, 45(5):378--383, 1999.

\bibitem{obataya1999c_acoustic}
E.~Obataya and M.~Norimoto.
\newblock {Acoustic properties of a reed (Arundo donax L.) used for the
  vibrating plate of a clarinet}.
\newblock {\em The Journal of the Acoustical Society of America},
  106:1106--1110, 1999.

\bibitem{lord2003viscoelasticity}
A.E. Lord.
\newblock {Viscoelasticity of the giant reed material Arundo donax}.
\newblock {\em Wood Science and Technology}, 37(3):177--188, 2003.

\bibitem{pinard2003musical}
F.~Pinard, B.~Laine, and H.~Vach.
\newblock {Musical quality assessment of clarinet reeds using optical
  holography}.
\newblock {\em The Journal of the Acoustical Society of America}, 113:1736,
  2003.

\bibitem{picart2007tracking}
P.~Picart, J.~Leval, F.~Piquet, J.P. Boileau, T.~Guimezanes, and J.P. Dalmont.
\newblock {Tracking high amplitude auto-oscillations with digital Fresnel
  holograms}.
\newblock {\em Optics Express}, 15(13):8263--8274, 2007.

\bibitem{picart2010study}
P.~Picart, J.~Leval, F.~Piquet, JP~Boileau, T.~Guimezanes, and JP~Dalmont.
\newblock {Study of the Mechanical Behaviour of a Clarinet Reed Under Forced
  and Auto-oscillations With Digital Fresnel Holography}.
\newblock {\em Strain}, 46(1):89--100, 2010.

\bibitem{mounier2008investigation}
D.~Mounier, P.~Picart, J.~Leval, F.~Piquet, J.P. Boileau, T.~Guimezanes, and
  J.P. Dalmont.
\newblock {Investigation of clarinet reed auto-oscillations with digital
  Fresnel holography}.
\newblock {\em The Journal of the Acoustical Society of America}, 123:3240,
  2008.

\bibitem{Guimezanes_2007}
T.~Guimezanes.
\newblock {\em Etude Expérimentale et Numérique de l'Anche de Clarinette}.
\newblock PhD thesis, Université du Maine, Le Mans, France, 2007.

\bibitem{casadonte1993perfect}
D.~Casadonte.
\newblock {The perfect clarinet reed? Vibrational modes of realistic clarinet
  reeds}.
\newblock {\em The Journal of the Acoustical Society of America}, 94:1807,
  1993.

\bibitem{casadonte1995}
D.~Casadonte.
\newblock {\em The clarinet reed: an introduction to its biology, chemistry and
  physics}.
\newblock PhD thesis, Ohio State University, 1995.

\bibitem{facchinetti2000application}
M.~Facchinetti, X.~Boutillon, and A.~Constantinescu.
\newblock {Application of modal analysis and synthesis of reed and pipe to
  numerical simulations of a clarinet}.
\newblock {\em The Journal of the Acoustical Society of America}, 108:2590,
  2000.

\bibitem{facchinetti2003numerical}
M.L. Facchinetti, X.~Boutillon, and A.~Constantinescu.
\newblock {Numerical and experimental modal analysis of the reed and pipe of a
  clarinet}.
\newblock {\em The Journal of the Acoustical Society of America}, 113:2874,
  2003.

\bibitem{jolliffe2002principal}
IT~Jolliffe.
\newblock {\em {Principal component analysis}}.
\newblock Springer verlag, 2002.

\bibitem{schnars1994direct}
U.~Schnars and W.~J{\"u}ptner.
\newblock {Direct recording of holograms by a CCD target and numerical
  reconstruction}.
\newblock {\em Applied Optics}, 33(2):179--181, 1994.

\bibitem{joud2009a_imaging}
F.~Joud, F.~Lalo{\"e}, M.~Atlan, J.~Hare, and M.~Gross.
\newblock {Imaging a vibrating object by Sideband Digital Holography}.
\newblock {\em Optics express}, 17(4):2774--2779, 2009.

\bibitem{joud2009bfringe}
F.~Joud, F.~Verpillat, F.~Lalo{\"e}, M.~Atlan, J.~Hare, and M.~Gross.
\newblock {Fringe-free holographic measurements of large-amplitude vibrations}.
\newblock {\em Optics Letters}, 34(23):3698--3700, 2009.

\bibitem{Claripatch}
{Claripatch SA website}.
\newblock \url{http://www.claripatch.com}.

\bibitem{WikiPCA}
{Wikipedia}.
\newblock \url{http://en.wikipedia.org/wiki/Principal_component_analysis}.

\bibitem{Systat}
{Systat}.
\newblock \url{http://www.systat.com}.

\bibitem{roylance2001engineering}
D.~Roylance.
\newblock {Engineering viscoelasticity in 3.11 OpenCourseWare }.
\newblock {\em Departament of Materials Science and Engineering Massachusetts
  Institute of Technology, Cambridge}, 2001.

\bibitem{chaigne2008acoustique}
A.~Chaigne, J.~Kergomard, and X.~Boutillon.
\newblock {\em {Acoustique des instruments de musique}}.
\newblock Belin, 2008.

\bibitem{dill2006continuum}
E.H. Dill.
\newblock {\em {Continuum mechanics: elasticity, plasticity, viscoelasticity}}.
\newblock CRC, 2006.

\bibitem{gaul2007experimental}
L.~Gaul and A.~Schmidt.
\newblock {Experimental Determination and Modeling of Material Damping}.
\newblock {\em VDI-Berichte; Schwingungsdämpfung}, 2003:17--40, 2007.

\bibitem{Heinrich_1991}
J.~M. Heinrich.
\newblock {Recherches sur les propriétés densitométriques du matériau cane de
  Provence et ses similaires étrangers; relation avec la qualité musicale;
  étude associée d'une mesure de dureté}.
\newblock Technical report, Ministère de la Culture, Direction de la Musique et
  de la Danse, France, 1991.

\bibitem{saltelli2008global}
A.~Saltelli, M.~Ratto, T.~Andres, F.~Campolongo, J.~Cariboni, and D.S. Gatelli.
\newblock {\em Global Sensitivity Analysis. The Primer}.
\newblock John Wiley \& Sons Chichester, England.

\end{thebibliography}

\end{document}